\documentclass[twocolumn,a4paper,pra,aps,10pt,superscriptaddress,floatfix]{revtex4-2}

\usepackage[utf8x]{inputenc} 
\usepackage{amsfonts}
\usepackage{amsmath}
\usepackage{graphicx} 
\usepackage{dcolumn} 
\usepackage{bm} 
\usepackage{braket} 
\usepackage{hyperref}
\usepackage[per-mode=symbol]{siunitx} 
\usepackage[capitalize]{cleveref}
\usepackage{lipsum} 
\usepackage{stmaryrd} 
\usepackage{booktabs}

\let\originalleft\left
\let\originalright\right
\renewcommand{\left}{\mathopen{}\mathclose\bgroup\originalleft}
\renewcommand{\right}{\aftergroup\egroup\originalright}

\DeclareSIUnit[quantity-product = ]\percent{\char`\%}

\DeclareMathOperator{\neigh}{N}

\newcommand{\qb}[2]{{\text{#1}\ensuremath{#2}}}
\newcommand{\qd}[1]{\qb{D}{#1}}
\newcommand{\qx}[1]{\qb{X}{#1}}
\newcommand{\qz}[1]{\qb{Z}{#1}}
\newcommand{\qa}[1]{\qb{A}{#1}}

\newcommand{\sigx}{\hat{X}}
\newcommand{\sigy}{\hat{Y}}
\newcommand{\sigz}{\hat{Z}}
\newcommand{\logical}{\mathrm{L}}
\newcommand{\XL}{\hat{X}_\logical{}}
\newcommand{\ZL}{\hat{Z}_\logical{}}

\newcommand{\CZ}{\textsf{CZ}}

\newcommand{\meas}[2]{M^{#1}_{#2}}
\newcommand{\stab}[2]{s^{#1}_{#2}}
\newcommand{\synd}[2]{\sigma^{#1}_{#2}}
\newcommand{\syndalt}[1]{\tilde\sigma_{#1}}

\newcommand{\B}{$\mathbf{B}$}
\newcommand{\SXZ}{$\mathbf{S}^{\sigx{}|\sigz{}}$}
\newcommand{\SY}{$\mathbf{S}^{\sigy{}}$}
\newcommand{\STX}{$\mathbf{ST}^{\sigx{}}$}
\newcommand{\STY}{$\mathbf{ST}^{\sigy{}}$}
\newcommand{\T}{$\mathbf{T}$}
\newcommand{\Tp}{$\mathbf{T}'$}
\newcommand{\HX}{$\mathbf{H}^{\sigx{}}$}
\newcommand{\HY}{$\mathbf{H}^{\sigy{}}$}
\newcommand{\MXY}{$\mathbf{M}^{\sigx{}\sigy{}}$}
\newcommand{\MZZ}{$\mathbf{M}^{\sigz{}\sigz{}}$}
\newcommand{\CC}{$\mathbf{C}$}
\newcommand{\mat}[1]{\bm{#1}}

\begin{document}

\date{\today}

\title{Experimentally Informed Decoding of Stabilizer Codes\\ Based on Syndrome Correlations}

\author{Ants~Remm}          \altaffiliation[Present address: ]{Atlantic Quantum, Cambridge, MA 02139, USA} \affiliation{Department of Physics, ETH Zurich, CH-8093 Zurich, Switzerland} \affiliation{Quantum Center, ETH Zurich, CH-8093 Zurich, Switzerland}
\author{Nathan~Lacroix}     \affiliation{Department of Physics, ETH Zurich, CH-8093 Zurich, Switzerland} \affiliation{Quantum Center, ETH Zurich, CH-8093 Zurich, Switzerland}
\author{Lukas~B\"{o}deker} \affiliation{Institute for Theoretical Nanoelectronics (PGI-2), Forschungszentrum J\"{u}lich, 52428 J\"{u}lich, Germany} \affiliation{Institute for Quantum Information, RWTH Aachen University, 52056 Aachen, Germany}
\author{Elie~Genois}        \affiliation{Institut Quantique, D\'{e}partement de Physique, Universit\'{e} de Sherbrooke, Sherbrooke J1K 2R1 QC, Canada}
\author{Christoph~Hellings} \affiliation{Department of Physics, ETH Zurich, CH-8093 Zurich, Switzerland} \affiliation{Quantum Center, ETH Zurich, CH-8093 Zurich, Switzerland}
\author{Fran\c{c}ois~Swiadek} \affiliation{Department of Physics, ETH Zurich, CH-8093 Zurich, Switzerland} \affiliation{Quantum Center, ETH Zurich, CH-8093 Zurich, Switzerland}
\author{Graham~J.~Norris}    \affiliation{Department of Physics, ETH Zurich, CH-8093 Zurich, Switzerland} \affiliation{Quantum Center, ETH Zurich, CH-8093 Zurich, Switzerland}
\author{Christopher~Eichler} \altaffiliation[Present address: ]{Department of Physics, Friedrich-Alexander University Erlangen-Nürnberg (FAU), Erlangen, Germany} \affiliation{Department of Physics, ETH Zurich, CH-8093 Zurich, Switzerland}
\author{Alexandre~Blais} \affiliation{Institut Quantique, D\'{e}partement de Physique, Universit\'{e} de Sherbrooke, Sherbrooke J1K 2R1 QC, Canada} \affiliation{Canadian Institute for Advanced Research, Toronto, Ontario M5G 1M1, Canada}
\author{Markus~M\"{u}ller} \affiliation{Institute for Theoretical Nanoelectronics (PGI-2), Forschungszentrum J\"{u}lich, 52428 J\"{u}lich, Germany} \affiliation{Institute for Quantum Information, RWTH Aachen University, 52056 Aachen, Germany}
\author{Sebastian~Krinner}\altaffiliation[Present address: ]{Zurich Instruments,  CH-8005 Zurich, Switzerland}  \affiliation{Department of Physics, ETH Zurich, CH-8093 Zurich, Switzerland} \affiliation{Quantum Center, ETH Zurich, CH-8093 Zurich, Switzerland}
\author{Andreas~Wallraff}    \affiliation{Department of Physics, ETH Zurich, CH-8093 Zurich, Switzerland} \affiliation{Quantum Center, ETH Zurich, CH-8093 Zurich, Switzerland} \affiliation{ETH Zurich - PSI Quantum Computing Hub, Paul Scherrer Institute, CH-5232 Villigen, Switzerland}

\begin{abstract}
    High-fidelity decoding of quantum error correction codes relies on an accurate experimental model of the physical errors occurring in the device.
    Because error probabilities can depend on the context of the applied operations, the error model is ideally calibrated using the same circuit as is used for the error correction experiment.
  Here, we present an experimental approach guided by a novel analytical formula to characterize the probability of independent errors using correlations in the syndrome data generated by executing the error correction circuit.
    Using the method on a distance-three surface code, we analyze error channels that flip an arbitrary number of syndrome elements, including Pauli $\sigy$ errors, hook errors, multi-qubit errors, and leakage, in addition to standard Pauli $\sigx$ and $\sigz$ errors.
    We use the method to find the optimal weights for a minimum-weight perfect matching decoder without relying on a theoretical error model.
    Additionally, we investigate whether improved knowledge of the Pauli $\sigy$ error channel, based on correlating the X- and Z-type error syndromes, can be exploited to enhance matching decoding.
    Furthermore, we find correlated errors that flip many syndrome elements over up-to-eight cycles, potentially caused by leakage of the data qubits out of the computational subspace.
    The presented method provides the tools for accurately calibrating a broad family of decoders, beyond the minimum-weight perfect matching decoder, without relying on prior knowledge of the error model.
\end{abstract}

\maketitle

\section{Introduction}
\label{sec:intro}

Recent performance advances in quantum computing with superconducting qubits~\cite{Kjaergaard2020a,Blais2021,Chen2021p,Krinner2022,Kim2023b, Acharya2024}, have enabled experimental demonstrations of complex quantum computations~\cite{Arute2019,Zhong2020a,Wu2021c}.
State-of-the-art devices, however, still lack the performance to solve real-world problems~\cite{Tazhigulov2022}.
Quantum error correction~(QEC) promises to exponentially reduce the effective logical error rate at the cost of a polynomial increase in the number of qubits~\cite{Shor1995,Knill1998,Terhal2015n}, as recently demonstrated experimentally~\cite{Acharya2024, Lacroix2024, Paetznick2024}.
A widely pursued route to achieve this error suppression is the use of stabilizer codes~\cite{Gottesman1997}, which rely on repeated measurements of a set of mutually commuting stabilizer operators.
Physical errors are accompanied by changes in the stabilizer values, called the syndrome. The change of logical operator values can be decoded from the syndromes.
While a main challenge in implementing quantum error corrected circuits remains the realization of large-scale devices with low physical error rates, a topic of rising relevance is the fast and accurate decoding of error syndromes extracted from large circuits~\cite{Battistel2023}.
Decoders for experimental quantum error correction data have so far been calibrated by numerically optimizing for maximal logical fidelity~\cite{Sundaresan2022,Sivak2024}, analyzing the correlations between syndrome elements~\cite{Wu2021c,Chen2021p,Krinner2022}, or by numerically optimizing to match the higher-order correlations between syndrome elements~\cite{Chen2021r,Acharya2023}.
The most performant decoders often rely either on an accurate error model of the device, or a large dataset of training data in the case of machine-learning-based decoders.

In this work, we focus on developing an accurate error model by analyzing correlations in the experimental syndrome data of the target error-correction circuit without relying on a theoretical device error model or on conducting separate calibration experiments.
We present an analytical method to calculate the probability of any independent error event that has a unique signature in the stabilizer measurement outcomes, based on the higher-order correlations in the experimental syndrome data.
The method allows for the characterization of the full error model using the same circuit as is used for the QEC experiment.
This error model can be used to optimize the parameters of various decoding algorithms, which we illustrate by presenting a correlated minimum-weight perfect matching (MWPM) decoder as an example. Specifically, the probabilities of the different error events inferred with our method translate to the weights of the matching graph.
In addition, the full error model can be used for characterizing device performance under the same conditions as for the QEC experiment.
For example, crosstalk or control errors can be identified from a discrepancy between Pauli $\sigx$ and $\sigy$ error probabilities, since simple gate error models that include only dephasing and relaxation processes predict equal physical error probabilities for $\sigx$ and $\sigy$~\cite{Tomita2014surfNoise}.
Correlated errors that flip syndrome elements over multiple rounds, on the other hand, can be indicative of qubit leakage~\cite{Miao2023, Marques2023, lacroix2023fast} or error sources such as high-energy impact events~\cite{McEwen2021b} that are detrimental to the performance of the logical qubit.
We apply the tools we present to experimental data from a distance-three surface code logical state preservation experiment performed on the 17-qubit device first introduced in Ref.~\onlinecite{Krinner2022}.

\begin{figure*}
    \includegraphics[width=\textwidth]{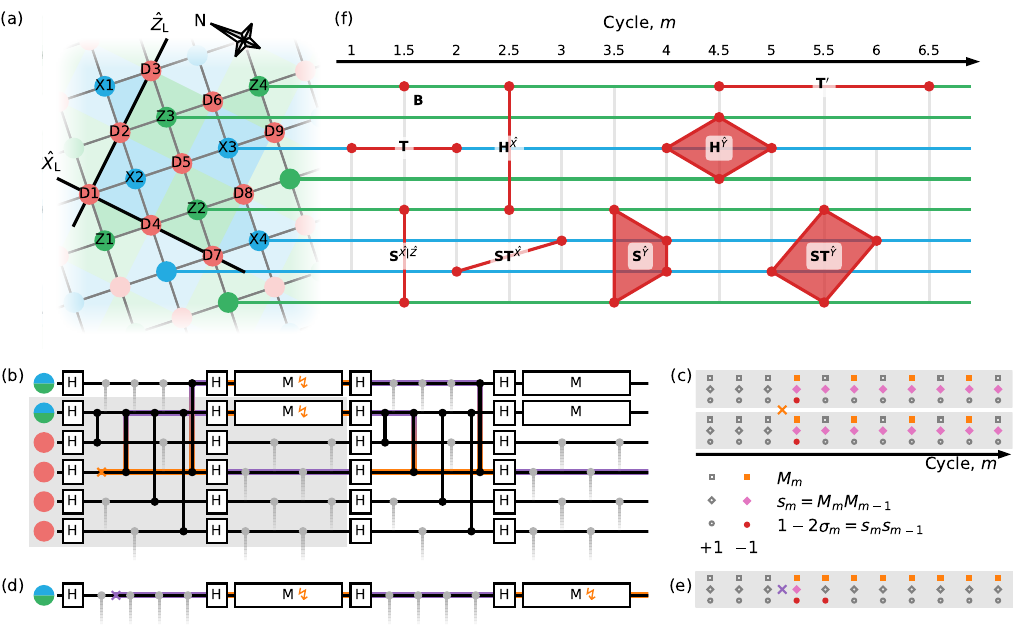}
    \caption{
        (a)~The surface code grid with data qubits in red and auxiliary qubits for Z- and X-type stabilizer measurement in green and blue, respectively.
        The physical device used in this work consists of the 17 qubits that are labeled.
        The support of the logical $\XL$ and $\ZL$ operators is indicated with black lines. A compass indicating the orientation of the lattice (pointing North) is used to define the East and West boundaries (see text).
        (b)~The circuit for two cycles of stabilizer measurements, with data qubits as red circles and auxiliary qubits as split blue-green circles.
        The circuit section highlighted with gray background implements the measurement of a single stabilizer, the building block of the surface code.
        The propagation of an $\sigx$ error (orange cross) is indicated with orange (as $\sigx$) and purple (as $\sigz$) lines.
        The readout outcomes that are flipped as a result of the $\sigx$ error are marked with $\lightning$.
        The grayed \CZ{} gates involve qubits that are not explicitly shown in this circuit.
        (c)~Conversion of the errors shown in (b) into stabilizer values $\stab{}{m}$ and syndrome elements $\synd{}{m}$.
        (d)~The propagation of a $\sigz$ error (purple cross) on an auxiliary qubit during the parity map.
        (e)~Conversion of the error in (d) into stabilizer values and syndrome elements.
        (f)~Examples of error signatures due to common single-qubit Pauli errors.
        Pauli $\sigx$ and $\sigz$ errors lead to two nonzero syndrome elements, categorized into time-like (\T), space-like (\SXZ), space-time (\STX), and hook (\HX) error classes.
        Pauli $\sigy$ errors lead to error signatures with support on up-to-four auxiliary qubits and classified into space-like (\SY), space-time (\STY), and hook (\HY) error classes.
        Errors on the boundary of the lattice can have a signature on a single auxiliary qubit only (\B{}) and readout misclassification errors lead to time-like signatures on syndrome elements two cycles apart (\Tp{}).
    } \label{fig:1}
\end{figure*}

The paper is structured as follows.
First, we introduce our implementation of the distance-three surface code and detail how we extract the syndromes for a logical state preservation experiment (\cref{sec:surface-code-syndromes}).
We then explain how we have implemented one of the most common decoding algorithms, the minimum-weight perfect matching decoder, in the experimental demonstration of quantum error correction with the surface code~\cite{Krinner2022} (\cref{sec:mwpm}).
Finally, we present the theoretical tools that allow us to characterize the error model of the device based on the experimental syndrome data (\cref{sec:error-extraction}).
We find a significant presence of errors that correlate with leakage of the data qubits out of the computational subspace and have signatures that span multiple syndrome elements over many cycles, which we discuss in detail in \cref{sec:diagnostics}.
These results highlight the capability of our method to determine the error probability associated with any observed syndrome signature and demonstrate its utility both in diagnosing device errors and in optimizing decoder parameters.

\section{From Errors to Syndromes}
\label{sec:surface-code-syndromes}

The experimental data we present was taken for a distance $d=3$ surface code~\cite{Krinner2022}.
We start by giving a brief description of the code and setting the terminology for the rest of the paper.
The surface code consists of a $d \times d$ square lattice of data qubits, see red circles in \cref{fig:1}~(a), which encode a protected quantum state.
As a stabilizer code~\cite{Gottesman1997}, it protects states that are simultaneous eigenstates of a set of commuting stabilizer generator operators (called \textit{stabilizers} here), each taking the value of~$\pm1$.
If we consider a square lattice with the data qubits at the vertices, the stabilizers of the surface code are the products of Pauli operators of the data qubits on the vertices of each plaquette.
The stabilizers $\hat{S}_\qa{i}$ alternate in the lattice between products of $\sigz$ and $\sigx$, see green and blue squares in \cref{fig:1}~(a),
\begin{equation}
    \label{eq:stabilizers}
    \hat{S}_\qx{i} = \prod_{\qd{j} \in \neigh(\qx{i})} \sigx_\qd{j}
    \quad \text{and} \quad
    \hat{S}_\qz{i} = \prod_{\qd{j} \in \neigh(\qz{i})} \sigz_\qd{j},
\end{equation}
where $\neigh(\qa{i})$ denotes the set of data qubits on the vertices of the stabilizer plaquette. 
At the center of each plaquette, an auxiliary qubit $\qa{i}=\qx{i}$ or $\qz{i}$ is used to measure the respective stabilizer.
Additionally, there are stabilizers located at the boundary of the surface code lattice with a support on only two data qubits. In total, the stabilizers of a distance $d$ surface code are adding up to $(d-1)^2 + 2(d-1) = d^2-1$ stabilizers.
By constraining the protected state to a mutual eigenstate of all the stabilizers, the dimension of the protected Hilbert space is reduced by a factor of two for each of the stabilizers, down to $2^{d^2} / 2^{d^2 - 1} = 2$, meaning that the protected subspace corresponds to a single \textit{logical qubit}.
The Pauli operators of the logical qubit are defined as two anticommuting operators
\begin{equation}
    \label{eq:logical-operators}
    \ZL = \prod_{\qd{j} \in \text{row}} \sigz_\qd{j}
    \quad \text{and} \quad
    \XL = \prod_{\qd{j} \in \text{column}} \sigx_\qd{j},
\end{equation}
which commute with all the stabilizers, see black lines in \cref{fig:1}~(a).

The stabilizers are repeatedly measured during the operation of the error correction code, yielding values $\stab{\qa{i}}{m}$ with $m$ the index of the error correction cycle and \qa{i} the associated auxiliary qubit.
In our implementation, we use the circuit shown in \cref{fig:1}~(b), which consists of single-qubit Hadamard gates implemented using $\pi/2$ rotations around $\sigy$ and virtual $\sigz$ rotations~\cite{McKay2017}, and four conditional phase flip~(\CZ{}) gates~\cite{Strauch2003, DiCarlo2009, Negirneac2021} that map the parity of the data qubits to the auxiliary qubits, which are thereafter read out~\cite{Fowler2012}.
To reduce the number of two-qubit gates which are executed in parallel, we apply the gates for Z-type stabilizers while reading out the X-type auxiliary qubits and \textit{vice versa}~\cite{Versluis2017,Krinner2022}.
Therefore, the stabilizers of different types are read out at full- and half-integer values of the cycle index $m=1, 1.5, 2,...$, respectively.
If the stabilizer value is $-1$ (indicating an odd parity), the circuit flips the state of the corresponding auxiliary qubit. Since we do not reset the auxiliary qubits between error correction cycles, the measurement outcome for an auxiliary qubit associated with a stabilizer value of $-1$ will alternate between $-1$ and $+1$ in consecutive cycles.
Therefore, we infer the stabilizer value $\stab{\qa{i}}{m} = \meas{\qa{i}}{m-1} \meas{\qa{i}}{m}$ from the change in consecutive readout outcomes $\meas{\qa{i}}{m} = \pm 1$ rather than by the measurement outcome in a given cycle.

Because the stabilizers involving $\sigx$ and $\sigz$ operators of each data qubit are repeatedly measured, all physical single-qubit errors are projected onto bit and phase flips~\cite{Greenbaum_2018,Bravyi2018}.
The bit and phase flips of the physical qubits will cause some stabilizers to flip and might also flip the logical operator values $\XL$ and $\ZL$, depending on where the errors occurred.
We express the change of the stabilizer values in terms of the \textit{syndrome elements} $\synd{\qa{i}}{m} = (1 - \stab{\qa{i}}{m-1} \stab{\qa{i}}{m})/2$, which have the value $\synd{\qa{i}}{m}=1$ or $0$, if the value of stabilizer $\hat{S}_\qa{i}$ at cycle~$m$, changed or did not change, respectively.
The process of deciding, based on the syndrome elements, whether the logical operator values have flipped, is called decoding.

To successfully decode errors, we must know which syndrome elements are flipped by each independent error process, i.e., the \textit{signature} of that error, and whether it flips any of the logical qubit Pauli operators.
As an example, let us consider a bit flip ($\sigx$) error on one of the data qubits, indicated by the orange cross in \cref{fig:1}~(b).
Note that this error is equivalent to a phase flip ($\sigz$) before the preceding Hadamard gate on that qubit.
The error propagates to two neighboring auxiliary qubits as phase flips via the \CZ{} gates, and the phase flips, in turn, change the outcomes of the following auxiliary qubit readouts.
As the effect of the error remains on the data qubit, the auxiliary qubits will be flipped in every consecutive cycle, leading to an alternating pattern of readout outcomes, see \cref{fig:1}~(c).
Converting the flips of readout outcomes $\meas{\qa{i}}{m}$ to flips of stabilizer values $\stab{\qa{i}}{m}$ and to flips of syndrome elements $\synd{\qa{i}}{m}$, we find that the signature of such an error includes two syndrome element flips.
The events are detected on neighboring auxiliary qubits during the same cycle.
We label all errors with such signatures \SXZ{} for errors with space-separated signatures due to $\sigx$ and $\sigz$ errors on data qubits.
Another common error is a phase flip ($\sigz$) error on an auxiliary qubit, see the purple cross in \cref{fig:1}~(d), or equivalently a bit flip error just before, during, or after readout.
In this case, the error flips all following readout outcomes $\meas{\qa{i}}{m}$ on that auxiliary qubit $\qa{i}$, which then results in two consecutive nonzero syndrome elements, see \cref{fig:1}~(e).
We label all errors with such signatures as \T{} for errors with time-separated signatures.

A few additional error classes are shown in \cref{fig:1}~(f), see \cref{sec:sup-syndrome-generation-examples} for examples of error propagation.
If an $\sigx$ error occurs on a data qubit between the two \CZ{} gates of one half-cycle, then the syndrome elements are flipped on the neighboring auxiliary qubits in consecutive cycles.
We label such errors as \STX{} for errors with a space-time separated signature.
If an $\sigx$ error occurs on an auxiliary qubit that is used to measure a weight-four Z-type stabilizer at the middle of the parity map, then this error propagates as phase flip errors to two neighboring data qubits which perform a gate with that auxiliary qubit next.
Depending on the gate order in the parity map, these phase flip errors will flip two or four neighboring syndrome elements of X-type.
We label such errors as \HX{} for $\sigx$-caused hook errors.
Bit flip errors on the auxiliary qubit at other times during the parity map lead to a phase flip error on zero, one, three, or four neighboring data qubits.
Because the neighboring qubits are part of a Z-type stabilizer, and the application of a stabilizer has no effect on the quantum state, then the error propagating to three or four neighboring qubits is equivalent to errors on the complementary one or zero neighboring data qubits.
Therefore, these errors are indistinguishable from phase flip errors on data qubits, which belong to the \SXZ{} class.
Again, due to the X-Z symmetry of the code, the situation is the same for $\sigx$ errors on X-type auxiliary qubits, in which case Z-type syndrome elements are flipped.

In addition to $\sigx$ and $\sigz$ errors discussed above, $\sigy$ errors can also occur at the same circuit locations.
These errors have signatures flipping both Z- and X-type syndrome elements.
We label the corresponding error classes as \SY{}, \STY{} and \HY{}.
Near the boundary of the surface code lattice, errors can have signatures that include only a single syndrome element, which we will label boundary errors \B{}.
Finally, we label readout misclassification errors, for which the auxiliary qubit state is incorrectly classified, but does not get flipped, as \Tp{} errors.
These errors are equivalent to a correlated bit flip just before and after a perfect measurement, and they flip two syndrome elements on a single auxiliary qubit $\Delta m = 2$ cycles apart.

We also consider two-qubit Pauli errors, not shown in \cref{fig:1}~(f), which might arise due to the two-qubit gates.
We distinguish between the \MZZ{} class, corresponding to a correlated phase flip error $\sigz \otimes \sigz$ on the two qubits involved in a two-qubit gate, and the \MXY{} class, corresponding to a correlated bit flip error $\sigx \otimes \sigx$, $\sigx \otimes \sigy$, $\sigy \otimes \sigx$, or $\sigy \otimes \sigy$.
Other two-qubit Pauli errors are equivalent to a single-qubit Pauli error on one of the qubits before or after the \CZ{} gate.
While we do not expect to observe correlated bit flips, errors of the \MZZ{} class could occur due to residual interactions between the qubits~\cite{Krinner2020} or a miscalibrated conditional phase of the \CZ{} gate.

\section{Minimum-Weight Perfect Matching Decoder}
\label{sec:mwpm}

There are various strategies for decoding the syndrome data in the surface code~\cite{duclos2010fast,herold2017cellular,Delfosse2022UFD,Sundaresan2022,delfosse2023splitting,higgott2023improved}.
Most accurate results can be achieved by maximum likelihood decoding~\cite{dennis2002topological}, in which case all possible combinations of physical errors that are consistent with the observed syndrome data are considered.
Each set of physical errors requires a corresponding correction of the logical qubit state, and the decoder picks the correction with the largest total likelihood.
Since the number of physical error combinations is exponentially large in code size and number of executed cycles, maximum likelihood decoding is prohibitively expensive for all but the very smallest of codes~\cite{Iyer2015}.
Approximations of likelihood calculation using tensor networks can reduce complexity~\cite{Bravyi2014}, but these methods remain too slow to meet the stringent timing requirements for large-scale error-corrected logical algorithms for which decoding must be performed between the application of non-Clifford gates in near-real time~\cite{Wu2021c,Acharya2023,Smith2023}.

When looking at one type of auxiliary qubits at a time, the signature of each physical error is given by at most two syndrome element flips.
Furthermore, a chain of neighboring errors leads to flipped syndrome elements only at the ends of the chain, since the syndrome element flips in the middle of the chain cancel with each other.
The resulting property that syndrome flips come in pairs, allows a much more efficient minimum-weight perfect matching~(MWPM) decoder~\cite{dennis2002topological,fowler2013minimum} to be used.
In MWPM decoding, all nonzero elements of the syndrome are matched in pairs.
Once the matching is completed, a unique logical correction operator can be inferred based on whether or not the logical operator is expected to be flipped by the error chains in the matched graph.
By assigning a weight to each potential pair of nonzero elements of the syndrome according to the likelihood of its occurrence, the problem of finding the most likely matching can be converted to finding the minimum weight matching.

Compared to maximum likelihood decoding, this leads to two approximations.
Foremost, by decoding the Z- and X-type stabilizers separately, we ignore correlations between syndrome types, which can contain extra information in the presence of errors that lead to flips of both types of stabilizers.
Second, we find the most likely matching, but ignore that several (each individually less likely) matchings can potentially lead to the same logical correction. Consequently, the logical correction with the highest probability of success might differ from the correction given by the most likely matching.

\begin{figure}[t]
    \includegraphics[width=\columnwidth]{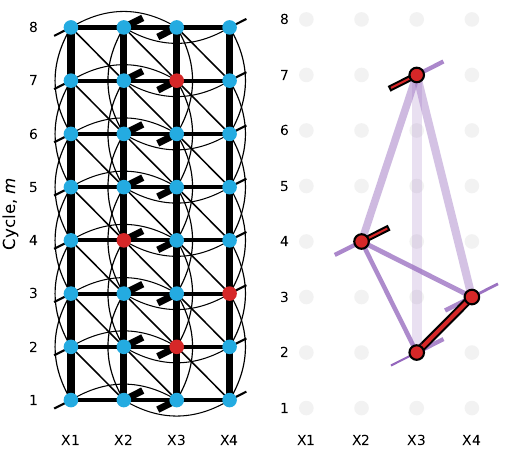}
    \caption{
        (a)~The auxiliary qubit graph for an example experimental run consisting of eight cycles of syndrome extraction.
        Vertices correspond to the measured syndrome elements (red if the syndrome element has a value of 1 and blue otherwise), while the edge thickness indicates the per-cycle probability of an independent error that flips the connected syndrome elements.
        The widths of the single-ended edges pointing to top-right and bottom-left indicate boundary error probabilities and their direction indicates to which boundary the syndrome element is connected.
        (b)~The syndrome graph.
        Vertices correspond to the nonzero syndrome elements of the auxiliary qubit graph, and the thickness and transparency of the edges indicate the weight for the MWPM decoding.
        The minimum-weight matching for this graph is indicated in red with a black border.
        Note that in the auxiliary qubit graph thick lines indicate probable errors, but in the syndrome graph, thin lines indicate low-weight error chains, i.e.,~probable error chains.
        Due to the small distance ($d=3$) of the code, the four X-type auxiliary qubits can be spacially connected in a one-dimensional chain through data qubits D2, D5, and D8 (see \cref{fig:1}(a)). This allows the corresponding syndrome graph to be visualized in a planar layout, with the horizontal axis representing the spatial dimension and the vertical axis representing different cycles.
        For larger code distances, the graph  would require a three-dimensional perspective.
    } \label{fig:2}
\end{figure}

Next, we will discuss the detailed procedure of minimum-weight perfect matching.
The first step, which can be done offline before running the surface code experiment, is to construct an \textit{auxiliary qubit graph} (also called a matching graph~\cite{Higgott2022}, ancilla graph~\cite{OBrien2017} or decoding graph), where each vertex $k$ corresponds to a syndrome element $\synd{\qa{i}}{m}$ of one type (X or Z), and each edge $q=(k,k')$ corresponds to a statistically independent and therefore uncorrelated error process, which flips the connected syndrome elements $k$ and $k'$, see \cref{fig:2}~(a).
In addition, there are two virtual vertices, which are connected to the syndrome elements that can flip without a pairing syndrome element at two opposing boundaries of the surface code lattice.
Physically, multiple distinct independent errors can have the same syndrome signature. Since we infer probabilities directly from the syndrome data, these errors are indistinguishable and treated as a single process to construct the auxiliary-qubit graph.
An error probability $p_q$ is associated with each edge of the auxiliary qubit graph.
The edges and their probabilities amount to the effective error model of the device, which can be constructed either from an independent physical error model or based on the correlations in the syndrome data~\cite{Spitz2018}, the latter of which is the main topic of this work, see \cref{sec:error-extraction}.

The second step of the MWPM decoding process is to construct the \textit{syndrome graph}, a fully connected graph where vertices correspond to nonzero syndrome elements of the auxiliary qubit graph for a given experimental run, see \cref{fig:2}~(b) for an example. 
Each edge $q$ in this graph is assigned a weight $w_q$ defined as the negative logarithm of the total probability that any chain of errors flips only the syndrome elements connected by that edge. 
Up to first order in the error probabilities, $w_q$ can be calculated as~\cite{OBrien2017}
\begin{equation}
    \label{eq:weight-calculation}
    w_q \approx - \ln \left( \sum_{R \in \mathcal{R}} \prod_{r \in R} p_r \right),
\end{equation}
where $\mathcal{R}$ denotes the set of possible paths between the endpoints of $q$ in the auxiliary qubit graph which do not go through the boundaries, and $r$ are the edges in one of those paths $R$.
A low edge weight indicates a high probability that a chain of errors triggers the corresponding syndrome pair, while a high weight suggests a low probability of such an event.
These weights can be precalculated once before the decoding, at the cost of only a polynomial overhead in the code distance, as~\cite{OBrien2017}
\begin{equation}
    \label{eq:weight-calculation-bulk}
    \mat{w} = - \ln \left( \left(\mat{1} - \mat{A}\right)^{-1} - \mat{1} \right),
\end{equation}
where $\mat{w}$ is a matrix of weights between all the potential nodes of the syndrome graph, $\mat{A}$ (\textit{adjacency matrix}) is the matrix of error probabilities in the auxiliary qubit graph, 
$\mat{1}$ is the identity matrix, and the logarithm is taken element-wise.

In a logical-state preservation experiment, where the same stabilizers are measured repeatedly, the system is expected to exhibit time-translational invariance. Consequently, the error probabilities are expected to remain constant from cycle to cycle during the middle of the error correction process, see \cref{sec:sup-synd-vs-cycle}.
Thus, one can calculate the error probabilities as a function of only the cycle difference $\Delta m = m' - m$ instead of cycles $m$ and $m'$,
\begin{equation}
    \label{eq:error-time-invariance}
    p_{(\synd{\qa{i}}{m}, \synd{\qa{j}}{m'})} = p^{(\Delta m)}_{\qa{i}, \qa{j}}.
\end{equation}
 
Note that in the first measurement cycles after the logical state preparation and in the last measurement cycles before the final measurement, the assumed time-translational invariance of the protocol does not hold. 
Therefore, using this approximation of time-invariant error probabilities can lead to slightly suboptimal performance.

For example, for X-type stabilizers in a $d=3$ surface code~\cite{Krinner2022}, we can calculate the adjacency matrix for a maximum cycle difference of $\Delta m_\mathrm{max} = 2$ as
\begin{equation}
    \label{eq:adjacency-matrix}
    \mat{A} =
    \begin{pmatrix}
        \mat{A}^{(0)}  & \mat{A}^{(1)}  & \mat{A}^{(2)}  & \mat{0}        & \mat{0}       & \mat{B} \\
        \mat{A}^{(-1)} & \mat{A}^{(0)}  & \mat{A}^{(1)}  & \mat{A}^{(2)}  & \mat{0}       & \mat{B} \\
        \mat{A}^{(-2)} & \mat{A}^{(-1)} & \mat{A}^{(0)}  & \mat{A}^{(1)}  & \mat{A}^{(2)} & \mat{B} \\
        \mat{0}        & \mat{A}^{(-2)} & \mat{A}^{(-1)} & \mat{A}^{(0)}  & \mat{A}^{(1)} & \mat{B} \\
        \mat{0}        & \mat{0}        & \mat{A}^{(-2)} & \mat{A}^{(-1)} & \mat{A}^{(0)} & \mat{B} \\
        \mat{0}        & \mat{0}        & \mat{0}        & \mat{0}        & \mat{0}       & \mat{0} \\
    \end{pmatrix},
\end{equation}
with the cycle-shifted adjacency matrices containg the probabilities of errors for which the signature spans $\Delta m$ cycles
\begin{equation}
    \label{eq:adjacency-matrix-cycle-shifted}
    \left(\mat{A}^{(\Delta m)}\right)_{i,j} = p^{(\Delta m)}_{\qx{i},\qx{j}},
    \quad \text{ and } \quad
    \left(\mat{A}^{(0)}\right)_{i,i} = 0
\end{equation}
and the boundary adjacency matrix contains the probabilities of errors with a signature on a single syndrome element
\begin{equation}
    \label{eq:adjacency-matrix-boundary}
    \mat{B} =
    \begin{pmatrix}
        p_{\qx{1},\qb{BE}{}} & 0                    \\
        0                    & p_{\qx{2},\qb{BW}{}} \\
        p_{\qx{3},\qb{BE}{}} & 0                    \\
        0                    & p_{\qx{4},\qb{BW}{}} \\
    \end{pmatrix}.
\end{equation}
Following the distance-three surface code layout and orientation indicated in \cref{fig:1}(a), we define an east and a west boundary (BE and BW). To construct the boundary adjacency matrix, note that
the stabilizers \qx{1} and \qx{3} connect only to the east boundary (\qb{BE}{} via \qd{3}, \qd{6}, and \qd{9}), and the stabilizers \qx{2} and \qx{4} to the west boundary (\qb{BW}{} via \qd{1}, \qd{4}, and \qd{7}). We have set the corresponding unconnected elements of $\mat{B}$ to zero.
We then apply \cref{eq:weight-calculation-bulk} to precalculate the weights for syndrome pairs up-to-$\Delta m_\mathrm{max}$ cycles apart.

As the final step, we run a minimum-weight perfect matching algorithm on the syndrome graph~\cite{Edmonds1965, OBrien2017, Higgott2022}.
The space-component of each matched edge corresponds to a set of data qubits where an error has occurred. 
 The time-component, arising from the space-time and auxiliary qubit errors, has no direct effect on the logical operator value.
When decoding the X-type syndrome, each overlap of a data qubit in a matched edge with the logical $\XL$ operator corresponds to a flip of $\XL$ that should be corrected.
Correspondingly, flips of the operator $\ZL$ can be inferred from the Z-type syndrome data.

Note that recent implementations of minimal-weight perfect-matching decoders~\cite{Higgott2023, Wu2023Fusion} achieve faster run-times by directly finding the perfect matching on the sparse auxiliary qubit graph. This approach eliminates the need to construct the syndrome graph with all-to-all connected edges and assigned weights~\cite{Higgott2023}.

With the goal of further improving the weight-inference-based MWPM decoding, we investigate the use of a correlated MWPM decoder (detailed in \cref{sec:appendix-decoding}). This decoder is designed to better correct for $\sigy$ errors, which flip both X-type and Z-type syndrome elements and thereby create correlations between the two syndrome types. For the current distance and error rates of the device, we do not observe significant performance improvements, suggesting that the logical error per cycle is not currently limited by $\sigy$ errors. Nevertheless, we expect this approach to become increasingly beneficial at larger distances~\cite{Paler2023pipelinedcorrelated, Bausch2024}, see~\cref{sec:appendix-decoding} for details.

\section{From Syndrome Correlations to Error Probabilities}
\label{sec:error-extraction}

Having a good quantitative knowledge of the experimental errors occuring on a device is of high importance for high-fidelity decoding, independently of the exact decoding algorithm used.
While many error processes can be characterized using independent measurements, the effective error rates might differ when running the actual error correction experiment, e.g., due to time-drift of parameters or unaccounted-for error mechanisms, like crosstalk.
Therefore, we ideally want to construct an error model for the decoder based on syndrome data produced by running the same circuit as for the error correction experiment.
In this section, we explain, based on experimental data, how this can be done.

Using the device presented in Ref.~\onlinecite{Krinner2022} (see also \cref{sec:sup-s17-device}), we prepare the logical state $\ket{0}_\logical$, $\ket{1}_\logical$, $\ket{+}_\logical$ or $\ket{-}_\logical$, acquire 16 cycles of stabilizer measurements, and finally read out all qubits. 
For each state, we perform 500,000 experimental runs, and remove the ones in which any of the qubits are measured outside the computational subspace, or the initial state is not the ground state, leaving us with about 54,000 runs per state. 
In the future, leakage reduction units as well as leakage-aware decoders could be employed to eliminate the need for post-selection on no-leakage events~\cite{Fowler2013,varbanov2020, Marques2023,lacroix2023fast, Miao2023}.
For calculating the syndrome elements for the first and last cycles, we make use of the known initial state of the auxiliary qubits, the initial parity of the data qubits, and the parity of the data qubits from the final readout.

The average per-cycle probability of detecting a non-zero syndrome element on weight-four and weight-two stabilizers is $\langle \synd{}{} \rangle = \num{0.165\pm0.017}$ and \num{0.118\pm0.006}, respectively, where the uncertainty indicates the standard deviation across different auxiliary qubits, see \cref{sec:sup-synd-vs-cycle}.
To visualize the correlations between syndrome elements, characteristic of the error classes discussed in \cref{sec:surface-code-syndromes}, we calculate the covariances $C^{(\Delta m)}_{\qa{i},\qa{j}}$ between syndrome elements on auxiliary qubits \qa{i} and \qa{j}, $\Delta m$ cycles apart as
\begin{equation}
    \label{eq:covariance}
    C^{(\Delta m)}_{\qa{i},\qa{j}} = \left\langle \synd{\qa{i}}{m} \synd{\qa{j}}{m+\Delta m} \right\rangle - \left\langle \synd{\qa{i}}{m} \right\rangle \left\langle \synd{\qa{j}}{m+\Delta m} \right\rangle.
\end{equation}
The averaging $\langle\cdot\rangle$ is done first over the experimental runs, and then over the cycle index~$m$.
We omit syndrome elements from the first and the last cycle when calculating the correlations, since these syndrome elements are measured based on the initial or final state of the data qubits, and are not representative of the mean value during the bulk of the experiment.

\begin{figure}
    \includegraphics[width=\columnwidth]{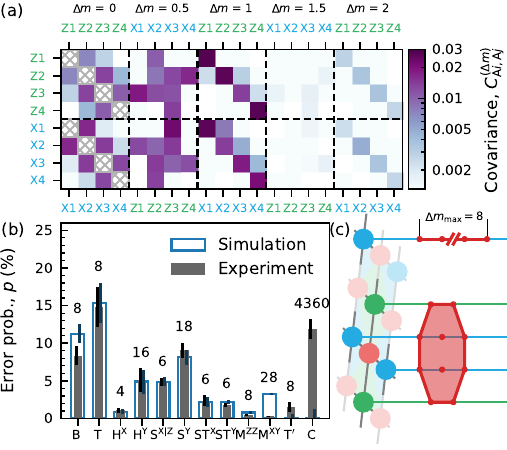}
    \caption{
        (a)~Average (over 16 cycles) covariance between pairs of syndrome elements as a function of the time-separation $\Delta m$ between them. The syndrome indicated by the column is detected $\Delta m$ cycles after the syndrome indicated by the row. The covarying qubit is indicated on the top or bottom axis for Z-type and X-type auxiliary qubits, respectively.
        (b)~Total error probability $p$ of the various error classes extracted from the experimental (solid gray bar) and simulated (blue wireframe) syndrome correlations. The number above the bar indicates the number of different error signatures that were considered in that class. The error bars indicate the standard deviation of the total, calculated as the square root of the sum of the squared deviations from the mean within that class.
        (c)~Schematic of the signatures that are considered as the highly-correlated \CC{} error class.
        That is, any subset of the highlighted syndrome elements on one auxiliary qubit over nine cycles or on multiple auxiliary qubits neighboring one data qubit separated by up to two cycles.
    } \label{fig:3}
\end{figure}

For $\Delta m = 0$, we observe the highest covariance between neighboring auxiliary qubits, see the first off-diagonal elements of the matrix in \cref{fig:3}~(a). These correspond to correlated syndrome element flips due to \SXZ{} and \SY{} errors.
The second off-diagonal, which has significant but lower covariance, corresponds to correlated syndrome element flips on next-nearest neighbors, caused by hook errors such as \HX{} and \HY{}.
We find negligible covariance between syndrome elements that are spatially separated by more than two data qubits (third off-diagonal), suggesting that errors remain local, as expected for uncorrelated single-qubit errors.
Because we use a pipelined stabilizer measurement circuit in which stabilizers of different types are measured sequentially, the covariance between syndrome elements of different types of auxiliary qubits is detected at a half-cycle separation, i.e.\ $\Delta m = 0.5$.
The nonzero covariances occur between neighboring auxiliary qubits and are caused by \SY{} and \STY{} errors.
We observe the highest covariance between syndrome elements on a single auxiliary qubit between consecutive cycles of error correction $\Delta m = 1$, which corresponds to \T{} errors.
The high covariance is an indication of the relatively higher probability of \T{}-class errors, being caused by auxiliary qubit dephasing and readout errors.
The nonzero covariances on the first off-diagonal are caused by \STX{} and \STY{} errors.
There is almost no covariance for $\Delta m = 1.5$, and only diagonal elements for $\Delta m = 2$.
An expected source of errors leading to $\Delta m = 2$ correlations is readout misclassification errors \Tp{}, where the auxiliary qubit state is misclassified without changing the qubit state.
However, as we discuss in \cref{sec:diagnostics}, most of these correlations are due to errors that can flip several syndrome elements over many cycles, possibly related to leakage of the data qubits outside the computational subspace.

For calculating the weights for minimum-weight perfect matching decoding, we have to convert those covariances into per-cycle error probabilities.
An analytical formula for this, assuming that every error flips at most two syndrome elements, is presented by Spitz \textit{et al.}~\cite{Spitz2018}.
Here, we have generalized those equations for errors with signatures that flip an arbitrary number of syndrome elements.
In the general case, the probability $p_{i_1,\dots,i_n}$, that an error that flips $n$ syndrome elements $\synd{}{i_1}, \dots, \synd{}{i_n}$ occurs, can be calculated as
\begin{equation}
    \label{eq:corr-to-perr-general}
    p_{i_1,\dots,i_n} = \frac{1}{2} - \frac{1}{2}\frac{
        \displaystyle\prod_{\begin{subarray}{r}
                \{j_1,\dots,j_m\} \subseteq \\ \;\; \{{i_1,\dots,i_n}\}
            \end{subarray}}
        \left\langle \syndalt{j_1} \dots \syndalt{j_m} \right\rangle^{(-1)^{m-1} 2^{-(n-1)}}
    }{
        \displaystyle\prod_{\begin{subarray}{r}
                \{j_1,\dots,j_m\} \supset \\ \;\; \{{i_1,...,i_n}\}
            \end{subarray}} \left(1 - 2 p_{j_1,\dots,j_m}\right)
    },
\end{equation}
see \cref{sec:sup-correlations-to-cycle-errors} for the verification of this formula and how it relates to the formulas introduced in Ref.~\onlinecite{Spitz2018}. 
Here, we have denoted $\syndalt{i} = 1-2\synd{}{i}$ for brevity, and the indices $i$ and $j$ each include both the auxiliary qubit index and the cycle number at which the syndrome element is measured.
The product in the numerator is taken over all the subsets $\{j_1,\dots,j_m\}$ of arbitrary length $m$ of the set of syndrome elements $\{i_1,\dots,i_n\}$, including the set itself, whereas the product in the denominator is taken over the supersets of a different arbitrary length $m$. We provide the explicit form of the equation when accounting for errors that can trigger up to four syndrome elements simultaneously in \cref{sec:sup-correlations-to-cycle-errors}.

Next, using \cref{eq:corr-to-perr-general}, we calculate the probabilities of errors that trigger various error signatures in the same dataset as was used to calculate the correlations shown in \cref{fig:3}~(a).
Because each run of the $N=16$-cycle-long experiment produces $(d^2-1)(2N-1)/2 = 124$ syndrome elements, it is unfeasible to calculate the error probability for all of the $2^{124}$ signatures.
We therefore only consider error signatures that are caused by single-qubit Pauli errors at any position in the circuit or two-qubit Pauli errors during a two-qubit gate, as presented in \cref{sec:surface-code-syndromes}.
To avoid introducing a bias into the extracted error probabilities by omitting the renormalization by $p_{j_1,\dots,j_m}$ in the denominator in \cref{eq:corr-to-perr-general}, we need to make sure that we include all processes with highly correlated signatures present in the system, see \cref{sec:sup-cycle-error-bias}.
Failure to do so may result in nonphysical (negative) error probabilities beyond statistical fluctuations that require additional corrections~\cite{Ali2024}.
To address this, we include two additional error signatures in the analysis that account for highly correlated errors, which we label \CC{}, see \cref{fig:3}~(c).
First, we consider any subset of flipped syndrome elements on up to nine consecutive cycles on a single auxiliary qubit, and second, we consider any subset of syndrome elements on auxiliary qubits neighboring one data qubit, separated by up to two cycles.
The maximum cycle separation for both of these signature classes is limited by the computational power, as the number of signature subsets grows exponentially with the maximum cycle separation.
The \CC{} error class could be associated with undetected leakage of the data qubits to higher transmon states, as we will show in \cref{sec:diagnostics}.
We average the probabilities of errors with the same signature over the cycles (except the first and the last cycle) to obtain the average probabilities for the 116 error signatures stemming from Pauli errors in the circuit and 4360 error signatures belonging to the \CC{} class.
The total probability of an error per cycle from each of the twelve signature classes is shown in \cref{fig:3}~(b).

We compare the experimental results to a circuit-level Pauli-error simulation, shown as blue wireframes in \cref{fig:3}~(b).
The simulation assumes uniform error probabilities across the device, with the probability values based on independent calibration measurements, see \cref{sec:sup-pauli-simulation-model} for details.
Despite this simplification, we observe overall good agreement between the sum of error probabilities of each class obtained from simulation and those extracted from the experiment. The most notable difference is in the correlated bit flips due to \CZ{} gates, indicated by the \MXY{} error class.  These errors are included in the depolarizing error model of the simulation, but are not as pronounced in the experimental data.
In simulations with a depolarizing Pauli noise model, two-qubit gate errors have been found to contribute most significantly to the logical error probability~\cite{Fowler2012}.
However, if correlated bit flip errors are rare for the experimental implementation of the \CZ{} gate, as we observe in our data, the effect of two-qubit gate errors on the logical error probability in experimental realizations might be more akin to single qubit errors.
Note that the simulations do not include leakage and readout misclassification errors, which could explain some of the residual differences between simulations and experimental data.

The in-situ error characterization method presented here is especially useful for stabilizer codes, since it yields the probabilities of errors with a given signature, which is exactly the information needed for decoding, without resorting to simulation of the circuit.
Furthermore, the method allows us to analyze spurious correlations between syndrome elements not caused by known Pauli errors, and to identify whether they lead to highly correlated errors, which are known to be particularly harmful in quantum error correction~\cite{Aharonov2006,Jouzdani2014a}.

\section{Device Diagnostics Using Error Probabilities}
\label{sec:diagnostics}

\begin{figure}
    \includegraphics[width=\columnwidth]{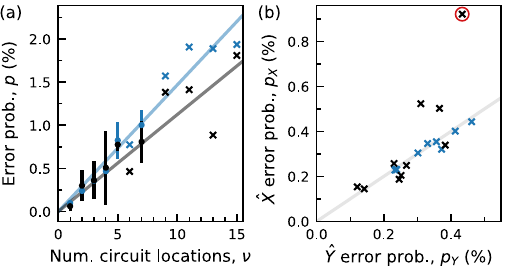}
    \caption{
        (a)~The probability $p$ of triggering an error signature as a function of the number of combinations of Pauli errors and circuit locations  $\nu$, that can cause the signature, extracted from experimental (black) and simulation (blue) data.
        (b)~Comparison of error probabilities, $p_X$ and $p_Y$, for signature pairs which can only be caused by a $\sigx{}$ or a $\sigy{}$ error, respectively, at a specific location.
        Black crosses correspond to experimental data and blue crosses to simulation.
        Due to the symmetry of the physical error mechanisms with respect to $\sigz$ rotations, we expect the two probabilities to be identical, which is indicated by the diagonal gray line. The red circle highlights error probabilities of \qb{D}{2} during the X-type parity map, which deviate significantly from the expectation (see text for details).
    } \label{fig:4}
\end{figure}

In this section, we study the per-cycle probabilities of specific errors with various signatures.
The signature of a circuit-level error in the syndromes depends on the Pauli error itself (for instance, single-qubit $\sigx{}$, $\sigy{}$, or $\sigz{}$ error) and on its location in the circuit. In addition, several combinations of Pauli errors and circuit locations can lead to the same error signature.
Therefore, we start by categorizing the error signatures according to the number $\nu$ of possible combinations of Pauli errors and circuit locations that can trigger it.
Specifically, we consider 15 two-qubit Pauli errors occurring at any of the 24 two-qubit gates of a single cycle, amounting to a total of $24 \cdot 15 = 360$ possible combinations, see \cref{sec:sup-full-circuit-diagram}.
We omit signatures from the time-like errors class \T{}  in this analysis, since readout errors strongly enhance their probability.
If uniform depolarizing noise at every two-qubit gate was the only error source, we would expect the error probability to be directly proportional to the number of combinations which can trigger that signature.
Overall, we indeed observe a strong correlation between the probability of an error signature and the number of combinations of Pauli errors and circuit locations that can produce it, see \cref{fig:4}~(a).
This correlation suggests that the variation in the probabilities of different error signatures is largely determined by how frequently errors that can produce each signature arise in the circuit.
The correlation is also reproduced by the simulation with independently characterized parameters [blue markers and line in \cref{fig:4}~(a)].

Another application of the generalized error probability extraction method is the identification of crosstalk and control errors.
For some circuit locations, there are unique signatures for both $\sigx$ and $\sigy$ errors, belonging to the classes of space-time errors or hook errors due to $\sigx$ or $\sigy$  type errors, denoted as \STX{} or \HX{} and \STY{} or \HY{}, respectively.
Since the physical energy relaxation and pure dephasing of the qubits is described by an error channel that is symmetric with respect to $\sigx$ and $\sigy$, we expect the corresponding extracted error probabilities $p_X$ and $p_Y$ to be equal~\cite{Tomita2014surfNoise}.
We find $p_X \approx p_Y$ for all error pairs in simulation and in experiment, see \cref{fig:4}~(b), except for errors on \qb{D}{2} during the X-type parity map (black cross circled in red), for which $p_X \approx 2p_Y$.
Control errors arising from miscalibrations that cause systematic under- or overrotations could explain this discrepancy because they can increase the likelihood of $\sigx$ errors without increasing the probability of $\sigy$ errors, causing $p_X$ to be larger than $p_Y$.
An alternative source for this undesired rotation could be microwave crosstalk during the X-type parity map affecting \qb{D}{2}. However, this explanation appears less likely, as an independent characterization of microwave crosstalk indicates that \qb{D}{2} is not significantly affected by microwave pulses applied to the drive lines of other qubits on the device~\cite[Supplementary Inf.]{Krinner2022}.

\begin{figure}[t]
    \includegraphics[width=\columnwidth]{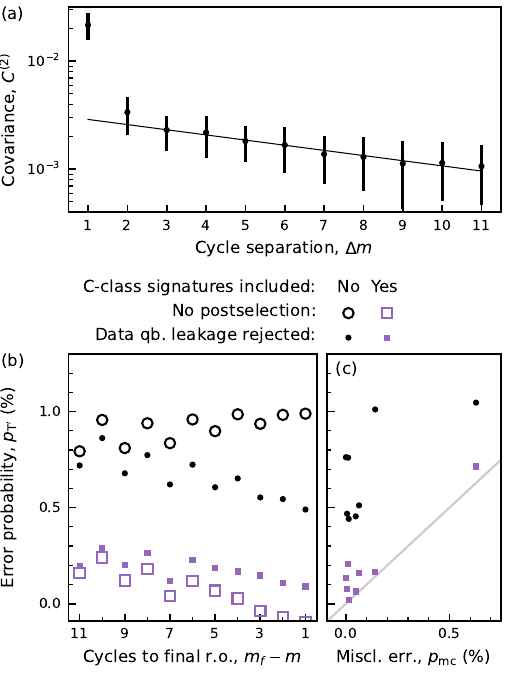}
    \caption{
        (a)~Covariance between syndrome elements on a single auxiliary qubit $\Delta m$ cycles apart, averaged over the eight auxiliary qubits.
        The error bars indicate one standard deviation over the different qubits.
        A fit to an exponential decay model for $\Delta m > 2$ is shown as a black line.
        (b)~Extracted probability of an error triggering a syndrome configuration corresponding to measurement misclassification \Tp{} ($\Delta m=2$) as a function of the number of cycles until the final data qubit readout.
        The four data series correspond to using data qubit leakage rejection or not and including leakage-related syndromes in the analysis or not.
        The data is averaged over the auxiliary qubits.
        (c) Average (over cycles) \Tp{} error probability of the eight auxiliary qubits as a function of the independently characterized readout misclassification error $p_\mathrm{mc}$.
        The gray line indicates the expected identity relation $p_{\mathbf{\mathrm{T}'}}=p_\mathrm{mc}$.
    } \label{fig:5}
\end{figure}

Next, we analyze the potential origin of the signature class of errors that cause longer time correlations, introduced as \CC{}.
As the first observation, we investigate the time-like correlations in the experimental syndrome data, and find that the covariance between syndrome elements on a single auxiliary qubit $\Delta m$ cycles apart decays exponentially as $0.89^{\Delta m}$, see \cref{fig:5}~(a).
Ideally, we expect the correlations for $\Delta m > 2$ to be zero.
The slow decay could be an indication of leakage of the data qubits~\cite{McEwen2021a} that is not removed by the leakage rejection scheme based on the final readout of the data qubits.
This hypothesis is supported by the observation that without including long-time-scale error-class signatures in the analysis, the error probability for \Tp{} signatures is higher at the beginning of the experiment when the leaked data qubits have several cycles to seep~\cite{Wood2017} back to the computational subspace before the final readout at the end of the experiment, see black filled dots in \cref{fig:5}~(b).
If we do not reject any runs based on the final data qubit readout, then the error probability does not depend on the cycle number and is comparable to the error probability with leakage rejection close to the first cycle, see open black circles in \cref{fig:5}~(b).
Alternatively, the slow decay could also be an indication of the probability of bit-flip errors varying during the data acquisition period, which would introduce correlations between errors with a large time-separation, see \cref{sec:sup-time-drift-example} for an example.
The varying error rate could be caused by the asymmetry of the energy relaxation channel and readout errors, implying that the effective bit flip error rate for auxiliary qubits depends on whether they spend more time in the ground or excited state, or by changes in the rate of quasiparticle generation and tunneling, for example due to impacts with cosmic rays~\cite{McEwen2021b}.
Because quasiparticle tunneling can also cause leakage~\cite{Serniak2018}, then it would be consistent with the suppression of errors when data qubit leakage rejection is used, but asymmetry of the energy relaxation channel alone could not explain our observations.

With the goal of taking into account highly correlated noise processes of the experiment, we include the \CC{} class signatures in the analysis.
Due to the renormalization term in the denominator of \cref{eq:corr-to-perr-general}, accounting for these additional processes gives us a more accurate value for the probability of \Tp{} errors.
With this analysis, we find that the error probability of readout misclassification errors \Tp{}  only weakly depends on the cycle number and whether data qubit leakage rejection is used or not, see filled and open purple squares in \cref{fig:5}~(b).
Furthermore, the error probabilities extracted using the improved analysis are consistent with the separately characterized overlap error of auxiliary qubit readout, which is the expected mechanism for \Tp{} errors, see \cref{fig:5}~(c).

\section{Conclusion}
\label{sec:conclusion}

An accurate error model of the quantum error correction circuit is a crucial component for any high-fidelity decoder.
In this work, we explained in detail how the syndrome is generated under a circuit-level Pauli noise model.
We presented a novel closed-form analytical method for calculating the error probabilities of errors with a given signature from the correlations between syndrome elements of arbitrary weight.
We used these error probabilities to calculate the weights of a minimum-weight perfect matching decoder as used in Ref.~\onlinecite{Krinner2022}, the nuances of which we also explained.
Furthermore, a correlated matching decoder is employed to harness the correlations introduced by $\sigy$ errors between both syndrome graphs to increase the decoding performance by a modest but stable amount.
In addition, the error model can be used to analyze the crosstalk, control, and leakage errors, measured using the same circuit as is used for executing the error correction experiment.
We identified control errors on one of the data qubits, leading to an imbalance of $\sigx$ and $\sigy$ errors.
We also identified errors with signatures spanning multiple syndrome extraction cycles, which are consistent with undetected leakage on the data qubits or quasiparticle tunneling.

Although characterizing errors based on syndrome data taken directly from executing the error-correction circuit can be a powerful tool, one needs to be aware of its limitations.
First, even though the analytical formula allows to easily calculate the probability of errors that trigger an arbitrary number of syndrome elements, it is not feasible to calculate the probabilities for triggering all possible signatures, as the number of signatures is exponentially large in the number of syndrome elements.
Therefore, some errors with non-standard signatures might be excluded from the model.
Second, this method requires many experimental runs of the error-correction circuit to accurately estimate the syndrome element correlations used to calculate the error probabilities.
Because of the renormalization terms in the denominator of \cref{eq:corr-to-perr-general}, the more high-weight signatures are included in the model, the more the uncertainties of errors with lower-weight signatures increase.
Furthermore, if the error model parameters drift during the time it takes to gather the necessary amount of statistics, then spurious correlations can appear in the data.

The main advantage of the presented method is that the error model is extracted from experimental data obtained by executing the same quantum circuit which we used to run the quantum error correction code.
This allows for the identification of error sources, such as crosstalk, that may not be captured in isolated characterization experiments.
Furthermore, the fact that the probability of errors with signatures of arbitrary weight can be calculated is beneficial for identifying error models which include $\sigy$ errors.
This can be used for improved decoders that account for correlations between X- and Z-type stabilizers, which we demonstrated for a modified version of the matching decoder~\cite{Fowler2013a}.
The inference of error processes causing higher-weight signatures can potentially benefit also other decoders, such as belief matching~\cite{higgott2023improved} and tensor-network-based~\cite{Bravyi2014} decoders.
As we have shown, the method can also be used to characterize errors with high-weight signatures, which can cause systematic deviations in extracted single- and two-qubit error probabilities if unaccounted for.

\section{Acknowledgements} \label{sec:acknowledgements}
The authors are grateful for valuable feedback on the
manuscript by N.~Delfosse, M.S.~Peralta, and P.~Girardin.

We acknowledge financial support by the Office of the Director of National Intelligence (ODNI), Intelligence Advanced Research Projects Activity (IARPA), through the U.S. Army Research Office grant W911NF-16-1-0071, by the EU Flagship on Quantum Technology H2020-FETFLAG-2018-03 project 820363 OpenSuperQ, by the National Center of Competence in Research `Quantum Science and Technology' (NCCR QSIT), a research instrument of the Swiss National Science Foundation (SNSF, grant number 51NF40-185902), by the SNSF R'Equip grant 206021-170731, by the EU programme H2020-FETOPEN project 828826 Quromorphic and by ETH Zurich.
S.K. acknowledges financial support from Fondation Jean-Jacques et F\'{e}licia Lopez-Loreta and the ETH Zurich Foundation.
L.B. and M.M. acknowledge support by the Deutsche Forschungsgemeinschaft through Grant No. 449905436. M.M. furthermore acknowledges support by the German Research Foundation (DFG) under Germany’s Excellence Strategy ‘Cluster of Excellence Matter and Light for Quantum Computing (ML4Q) EXC 2004/1’ 390534769, and by the ERC Starting Grant QNets through Grant No. 804247. L.B. gratefully acknowledges the computing time provided to us at the NHR Center NHR4CES at RWTH Aachen University (project number p0020074).
E.G. and A.B. acknowledge support from NSERC, the Canada First Research Excellence Fund, and the Minist\`ere de l’\'Economie et de l’Innovation du Qu\'ebec.
This is funded by the Federal Ministry of Education and Research, and the state governments participating on the basis of the resolutions of the GWK for national high performance computing at universities.

The views and conclusions contained herein are those of the authors and should not be interpreted as necessarily representing the official policies or endorsements, either expressed or implied, of the ODNI, IARPA, or the U.S. Government.

\begin{appendix}
    \section{Error Signatures}
    \label{sec:sup-syndrome-generation-examples}

\begin{figure*}[t]
        \includegraphics[width=\textwidth]{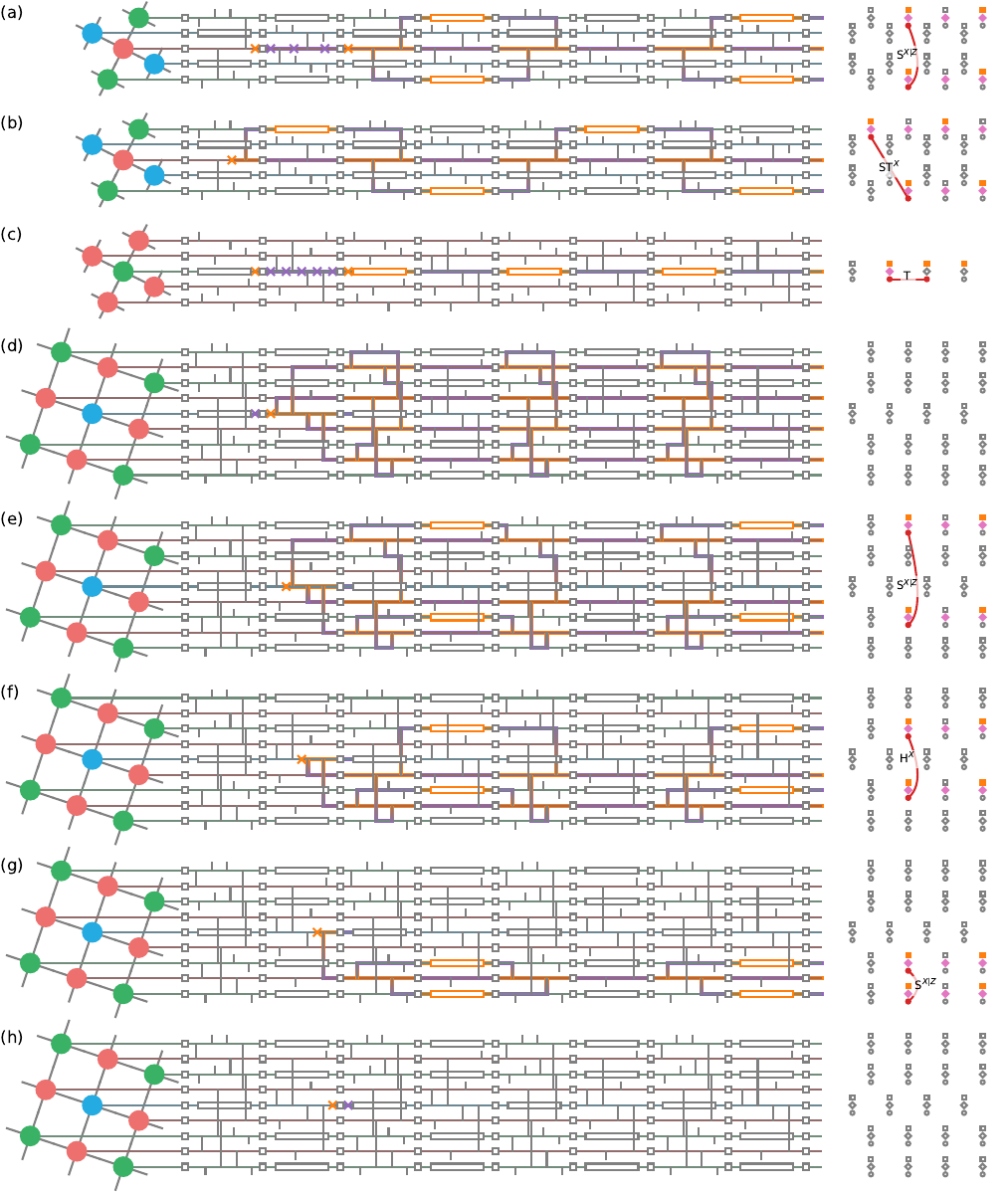}
        \caption{
            Examples of syndrome generation due to a physical $\sigx$ (orange cross) and $\sigz$ (purple cross) errors at all possible circuit locations. Multiple crosses in the same panel indicate equivalent errors.
            The squares indicate Hadamard gates, vertical lines \CZ{} gates and rectangles readouts. Orange and purple lines indicate the propagation of $\sigx$ and $\sigz$ flips through the circuit. At the right hand side of each panel, the readout, stabilizer and syndrome element flips are indicated in orange, pink, and purple, respectively. Errors on data qubits with signatures of class \SXZ{} and \STX{} are shown in panels (a) and (b), respectively. Errors on auxiliary qubits with \T{}, empty, \SXZ{}, \HX{}, \SXZ{}, and empty signature are shown in panels (c) to (h), respectively.
        } \label{fig:syndrome-all-examples}
    \end{figure*}

    To see which error signatures are expected due to physical errors, we place single-qubit $\sigx$ and $\sigz$ errors between each pair of neighboring operations in the syndrome extraction circuit, and propagate the errors through the circuit to see which auxiliary qubit readout outcomes are flipped due to the error.
    The set of single-qubit Pauli errors shown in~\cref{fig:syndrome-all-examples} covers all possible error locations if we account for the symmetry of the error extraction circuit with respect to change of the auxiliary qubit types.
    The propagation and signatures of single-qubit $\sigy$ errors and errors of class \MZZ{} and \MXY{} can be constructed as the sum of the propagations of the constituent $\sigx$ and $\sigz$ errors.

\section{Syndrome Probability vs Cycle}
    \label{sec:sup-synd-vs-cycle}

    \begin{figure}[t]
        \includegraphics[width=\linewidth]{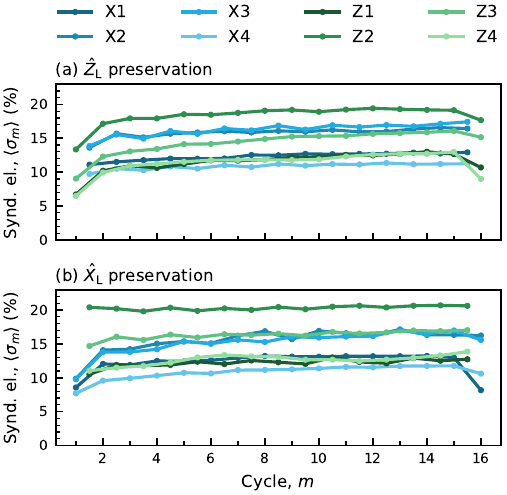}
        \caption{
            Mean syndrome element $\langle \synd{}{m} \rangle$ of each qubit as a function of the measurement cycle number $m$ when preparing an eigenstate of the (a) $\ZL$ and (b) $\XL$ operators.
        } \label{fig:syndrome-element-vs-cycle}
    \end{figure}

    To motivate the assumption of cycle-independent error rates, we calculate the mean syndrome element values for each stabilizer and each cycle, see \cref{fig:syndrome-element-vs-cycle}.
    The stabilizers matching the type of the prepared state are measured on whole-integer cycles, while the other stabilizers are read out on half-integer cycles in the piplined surface code circuit.
    The syndrome elements at the middle of the experiment are calculated based on auxiliary qubit measurements flip over two error correction cycles and are therefore sensitive to physical errors happening during two error correction cycles.
    On the other hand, the first and last syndrome elements for the stabilizer type of the prepared logical operator ($\ZL$ or $\XL$) make use of data qubit readout results to calculate the value of one of the stabilizers and is sensitive to errors during a single round of error correction.
    Therefore the syndrome elements and error probabilities at the time boundaries are expected to have a different mean value.
    Overall, we find that the mean syndrome element values are increasing very slightly over the course of the 16 experimental cycles.

    \section{Correlated MWPM Decoding}
    \label{sec:appendix-decoding}

    In this Appendix, we present a correlated MWPM decoder, with the goal of improving the weight-inference-based MWPM decoding described in the main text. We refer to the latter approach here as `standard' or uncorrelated MWPM. 
    The correlated MWPM decoder is designed to better correct for $\sigy$ errors, which flip both X-type and Z-type syndrome elements and thereby create correlations between the two syndrome types.
    The idea of harnessing the correlation of both syndrome types and performing a correlated decoding has been explored in complementary works such as \cite{Fowler2013a,delfosse2014decoding,yuan2022modified,higgott2023improved,Fujisaki2023isingDecoding,Paler2023pipelinedcorrelated,Delfosse2022UFD, delfosse2023splitting, cain2024correlated}.
    Our strategy for making use of the correlations between syndrome types is to iterate between decoding the two syndrome types, updating the weights based on the results from the other decoding graph.

    For syndrome matching, we can only utilize errors with signature weight of at most two.
    Including higher-order error signatures in the decoding graph would turn it into a hyper-graph (where edges can connect more than two vertices), on which finding the minimum matching is no longer efficient.    
    A $\sigy$ error can have a signature of weight two, three, or four when considering the joint syndrome, but not more than weight two on either syndrome type. 
    Therefore in the standard MWPM, the high-weight syndromes of $\sigy$ errors are split and a matching can be performed at the expense of treating single $\sigy$ errors as two uncorrelated $\sigx$ and $\sigz$ errors.
    In the correlated MWPM, we first decode one syndrome graph in the standard way. Using the information about the decoded errors, we update the other auxiliary qubit graph with the conditional error probabilities, given the decoded errors on the first graph.
    
    As an example, consider an error signature flipping three syndrome elements -- two Z-type and one X-type.
    Let us assume the X-type syndrome decoding is performed first with non-modified weights.
    Then the deduced correction will serve as a flag for the Z-type syndrome decoding to consider the conditional probability of having a $\sigy$ error given a $\sigz$ error instead of the joint $\sigx$ error probability.
    For the conditional probability, the leading-order contribution consists of taking the $\sigy$ error probability divided by the probability of finding the local X-type syndrome which was corrected in the previous Z-type syndrome decoding, 
    see \cref{eq:update_example}.
    
    Based on this procedure, we can alternate the decoding of both syndrome types while updating the weights until neither decoding outcome changes any more or until a predetermined maximum number of decoding rounds is performed.
    In practice, we only find minor changes to the decoded logical fidelity after the first iteration.
    We therefore restrict ourselves to one iteration of the weight update.
    In addition, to control the statistical uncertainty of the new weight, we don't fully replace the weight, but take a weighted average between the old and the new value instead.
    We find the highest logical fidelity for a weight of $\gamma = 0.09$, where $\gamma=0$ and $\gamma=1$ correspond to keeping the old weight and fully replacing the weight, respectively.
    This implies that for too large values of $\gamma$ the modification of the weights to include the possibility of $\sigy$ errors also breaks the correct decoding for some instances, see \cref{subsec:correlated-decoding-interpolation} for details.
    We suspect that statistical uncertainties of the inferred weights are at the origin of this problem.
    
\begin{figure}
        \centering
        \includegraphics[width=\columnwidth]{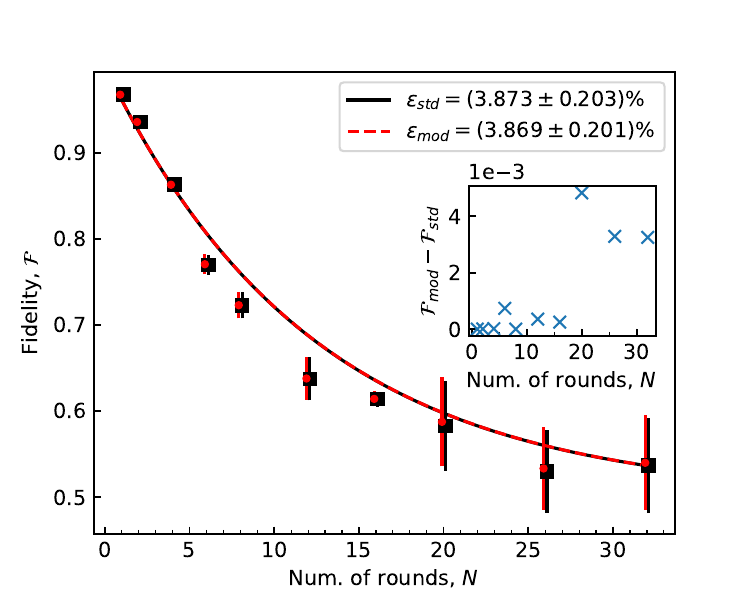}
        \caption{
            Comparison of the MWPM decoding fidelity of the $\ZL$ basis with initial state $\ket{0}_L$ using the standard weights (std, black) of syndrome weight up to two, not taking into account $\sigy$ errors and the modified weight (mod, red) where the decoding is repeated once and weights are updated using conditional probabilities based on the inferred $\sigy$ error signatures.
            For better visibility, the error bars are scaled by a factor 10.
            The inset shows the differences in fidelity of the two MWPM decoding variants. The employed weights were calculated based on the same data used for calculating the shown fidelities.
        }
        \label{fig:fidelity}
    \end{figure}

    Overall, we observe a slight improvement in decoding performance when using the correlated MWPM decoding protocol compared to the standard approach, although the difference is not statistically significant (see \cref{fig:fidelity}). We expect this method to deliver more substantial performance gains at larger code distances. In that regime, the decoder's enhanced ability to accurately account for $\sigy$ errors allows for more informed decisions among a broader set of correction options, thereby potentially reducing logical error rates. This expectation is supported by simulations~\cite{Paler2023pipelinedcorrelated} and has been experimentally validated in Ref.~\onlinecite{Bausch2024} using data from Ref.~\onlinecite{Acharya2023}.

In the subsections below, we provide further background on the principles of correlated MWPM using illustrative examples and elaborate on our implementation of the decoder.

    \subsection{Illustration of the probability update} \label{subsec:correlated-decoding-example}
    To give an example of how probabilities can be updated during the correlated matching, we consider a simplified setting where symmetric depolarising errors are placed on data qubits with probabilities $p^X=p^Z=p^Y=p/3$ before a perfect syndrome readout.
    Accordingly, all edges in both the X-type and the Z-type decoding subgraphs are given a weight of $w=-\ln{2p/3}$. 
    Now we assume that a physical $\sigy$ error occurred and that we perform the matching on the X-type syndrome subgraph first with the initial weights. 
    The decoding yields the correct result of applying $\sigz$ as a correction. 
    We can now reevaluate the probability of having had a $\sigx$ error for the Z-type syndrome decoding problem. 
    This new probability is given by the conditional probability of having $\sigx$ conditioned on having found $\sigz$ 
    \begin{equation}
    p(\sigx|\sigz)=\frac{p(\sigx \land \sigz)}{p(\sigz)}=\frac{p^Y+\mathcal{O}(p^2)}{p^Y+p^Z+\mathcal{O}(p^2)}=\frac{1}{2}+\mathcal{O}(p).
    \end{equation}
    Consequently, the probability of having an $\sigx$ error at the specific location that was heralded by the decoding of the X-type syndrome increases the probability from $2p/3$ to $\approx 1/2$. Here the value of $1/2$ is the coin flip probability between having found a genuine $\sigz$ error or a $\sigy$ error which in turn would carry the $\sigz$ error in addition to $\sigx$.
    As a next step the decoding of the Z-type syndrome can be performed on a graph where all edges but one carry the weight $w=-\ln{2p/3}$ and the edge that corresponds to a $\sigx$ correction has a weight of $w=-\ln{1/2}$, due to the adapted probability.

    \subsection{Correlated decoding under circuit level noise} \label{subsec:correlated-decoding-circuit-noise}
    For circuit-level noise and inferred error probabilities, we update a dressed probability by replacing it with the conditional probability for having an effective $\sigx$ ($\sigz$) error. This update is conditioned on finding the complementary effective $\sigz$ ($\sigx$) error from the previous decoding outcome.
    The concrete update signature and conditional probabilities depend on the circuitry and how $\sigy$ errors propagate.
    To illustrate this with an example, we assume circuit-level $\sigy$ errors with probability $p^Y_{(\synd{Z,\qa{i}}{m}, \synd{Z,\qa{j}}{m'}, \synd{X,\qa{k}}{m''})}$ that trigger the Z-type stabilizers $(A_i,A_j)$ at times $(m,m')$ and the X-type stabilizer $A_k$ at time $m''$.
    These elementary errors are additional to genuine $\sigx$ and $\sigz$ errors that have a signature one syndrome type.
    For simplicity, we are going to use the notation $\alpha=(A_i,m)$ to indicate the space-time coordinate of a syndrome element.
    We denominate the corresponding elementary error as $\sigy_v$ which causes the syndrome signature $v=(\synd{Z}{\alpha}, \synd{Z}{\beta}, \synd{X}{\gamma})$ and occurs by probability $p^Y_{(\synd{Z}{\alpha}, \synd{Z}{\beta}, \synd{X}{\gamma})}$.
    We assume now a setting where we perform an initial uncorrelated MWPM decoding on the X-type syndrome graph by which the matching suggests that a $\sigz_{v}$ error has happened.
    Such interpretation between matching and error suggestion relies on the notion of elementary errors. For the latter we assume that an edge which is included in the matching attributes a most likely error on the data qubits that can be taken as the correction.
    Considering the concrete matching example, the flip of the single syndrome element $\synd{X}{\gamma}$ is matched to a boundary, which corresponds to an $\sigz$ error at a location specified by $v$.
    Having obtained this information, we would like to properly adapt the error probability $\Tilde{p}^X_{(\synd{Z}{\alpha}, \synd{Z}{\beta})}$ of a $\sigx$ type likeliest error that is attributed to the edge $(\alpha,\beta)$. The weight of this edge would be updated for a round of correlated bit flip decoding.
    This is done by replacing $\Tilde{p}^X_{(\synd{Z}{\alpha}, \synd{Z}{\beta})}$ with the conditional probability of finding a $\sigx_{v}$ error given a $\sigz_{v}$ error is present (known from decoding)
    \begin{equation}
        p(\sigx_{v} | \sigz_{v}) = \frac{p(\sigx_{v} \land \sigz_{v})}{p(\sigz_{v})} \approx  \frac{ p^Y_{(\synd{Z}{\alpha}, \synd{Z}{\beta}, \synd{X}{\gamma})}}{\sum_{\alpha',\beta'}p^Y_{(\synd{Z}{\alpha'}, \synd{Z}{\beta'}, \synd{X}{\gamma})}+\Tilde{p}^{Z}_{\synd{X}{\gamma}}}.
        \label{eq:update_example}
    \end{equation}

    It is important to stress that this update only takes place if and only if the matching correction, here $\sigz_{v}$, is suggested by the complementary decoder.
    To explain the conditional probability of~\cref{eq:update_example}, firstly we note that it is a leading order approximation in terms of the error rate.
    The numerator is given by the probability of having both an $\sigx$ and a $\sigz$ error at the appropriate spacetime coordinate.
    To leading order, this corresponds to the inferred $\sigy$ error probability.
    In the next order terms like $p^X \cdot p^Z$ would contribute as well but are omitted in this work.
    In the denominator of~\cref{eq:update_example} we find the leading order probability of exhibiting an $\sigz_{a_i}$ error, which can emerge due to a proper $\sigz$ error with probability $\Tilde{p}^{Z}_{\synd{X}{\gamma}}$ or by means of a $\sigy$ error.
    With regard to the latter, there might exist a set of $\sigy$ errors which are degenerate with respect to the signature on the X-type syndrome. That is, there are multiple $\sigy$ errors that cause the same X-type syndrome but different Z-type syndromes.
    Over the corresponding probabilities, we perform a summation $\sum_{\alpha',\beta'}$.
    The adaption of the weight in this manner can be done for the Z-type syndrome decoding as well given a previous X-type syndrome decoding serving as the flag for it to be applied.
    The weight update in this way can also be applied for $\sigy$ error propagations that have weight-two or weight-four signatures on the joint syndrome.
    The correct weight update depends on these propagations which in turn are dictated by the stabilizer readout circuitry.\\

    \subsection{Application to the experimental data and interpolation approach} \label{subsec:correlated-decoding-interpolation}
\begin{figure}[t]
        \includegraphics[trim={3cm 0 2.5cm 0},clip,width=1\columnwidth]{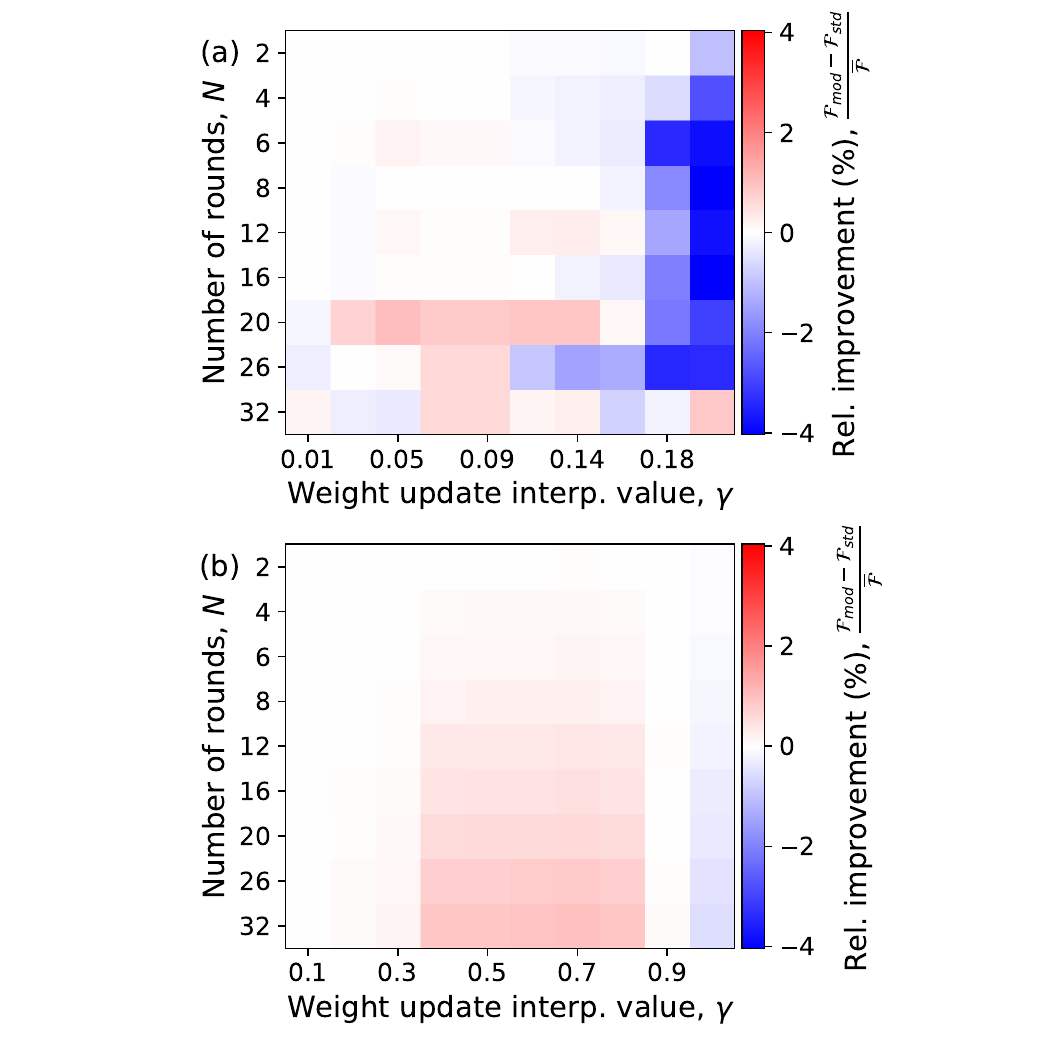}
        \caption{
            The relative improvement in fidelity of repeating the MWPM once after updating the weights as a function of the number of error correction cycles $N$ and the update strength $\gamma$ for (a) the experimental data and (b) simulated data under depolarizing circuit level noise. The relative improvement is defined as $\frac{\mathcal{F}_{mod}-\mathcal{F}_{std}}{\overline{\mathcal{F}}}$, where $\overline{\mathcal{F}}=\frac{\mathcal{F}_{mod}+\mathcal{F}_{std}}{2}$ is the average fidelity.
        }
        \label{fig:both_matrix}
    \end{figure}

    We perform weight updates that are compatible with the syndrome signatures of all possible $\sigy$ error propagations based on the circuitry in~\cref{sec:sup-full-circuit-diagram}.
    Note that the syndrome according to a $\sigy$ error propagation needs not to be contained in one measurement cycle but can also span between two consecutive rounds of stabilizer readout depending on the gate order.
    As mentioned in the main text, given an update rule, the decoding of both sub-graphs can be repeated until no change in the decoding is observed.
    Practically, it is sufficient to only repeat the decoding once, taking the modified weights for the experimental data analyzed in this work.
    The conditional probability, used for modifying the error probabilities, comes with a statistical error as it is a function of quantities inferred from the experimentally observed syndrome statistics.
    This uncertainty is genuinely larger than the uncertainty of the unaltered probability.
    We observe that this can have a deteriorating effect on the decoding result.
    Therefore, we relax the update of the probability by interpolating between the old and the new error probability.
    For the previous example update, that means to replace the standard weight by
    \begin{align}
        w'_{(\synd{Z}{\alpha}, \synd{Z}{\beta})}= & -(1-\gamma)\ln(\Tilde{p}^X_{(\synd{Z}{\alpha}, \synd{Z}{\beta})})                                                                                                                                                          \\
                                                  & -\gamma \ln\left(\frac{ p^Y_{(\synd{Z}{\alpha}, \synd{Z}{\beta}, \synd{X}{\gamma})}}{\sum_{\alpha',\beta'}p^Y_{(\synd{Z}{\alpha'}, \synd{Z}{\beta'}, \synd{X}{\gamma})}+\Tilde{p}^{Z}_{\synd{X}{\gamma}}}\right)\nonumber,
    \end{align}
    where $\gamma\in[0,1]$ is an interpolation parameter which we set to $0.09$ for the decoding results shown in \cref{fig:fidelity}.
    We observe that for larger values of $\gamma$ the fidelity is not constantly improved for each number of measurement rounds, see \cref{fig:both_matrix} (a).
    Already for values slightly above $\gamma=0.09$ the decoding performance worsens for small and larger number of rounds.
    For only few syndrome readouts, this decreased fidelity persists also for larger values of $\gamma$.
    As this maximal $\gamma$ value for obtaining a good performance in decoding is relatively low, we conjecture that at this level of statistical fluctuations of the weights, the influence of these fluctuations outweigh the improvement of including $\sigy$ error information such that minimizing the uncertainty is prioritized to improve the decoding.

    \subsection{Influence of statistical fluctuations} \label{subsec:correlated-decoding-fluxtuations}
    To investigate the effect of the statistical error on the decoding performance qualitatively, we re-run the modified and unmodified decoding on a simulated dataset with larger size and therefore lower statistical fluctuations compared with the experimental data.
    We perform simulations of the distance-three surface code with identical circuitry under depolarizing circuit-level noise.
    For this model, each circuit element exhibits a depolarizing error with probability $p$ after the execution of the circuit element.
    Of course, such simplistic one-parameter noise model is not intended to fully capture the actual experiment of Ref.~\onlinecite{Krinner2022} but shall only serve as a tool to understand the effect of statistical fluctuations of the weights on the decoding performance.
    We perform such simulation for a fixed error rate of $p=10^{-3}$ and a sample size that has about $12$ times more syndrome readouts as compared to the experimental data.
    These readouts distribute into $10^6$ shots per bin of number of readout rounds.
    Note that in the experimental data, especially for a larger number of readout rounds, the data sets are much smaller.

    We expect a lowered statistical uncertainty for this numerically generated dataset compared with the experimental data, and we infer weights from the sampled data and perform both the uncorrelated and the modified MWPM.
    We indeed find that now much larger values of $\gamma$ correspond to a consistent performance increase.
    We find that this improvement of the modified over the uncorrelated MWPM decoder only breaks down for values as large as $\gamma> 0.8$, see \cref{fig:both_matrix} (b), instead of $\gamma>0.09$ as for the experimental data.
    Another intriguing difference for the behavior of the relative improvement comparing experimental and simulated data is the smoothness that can be observed in the latter.
    
\begin{figure}
        \includegraphics[trim={0cm 0cm 0cm 1cm}, clip, width=\columnwidth]{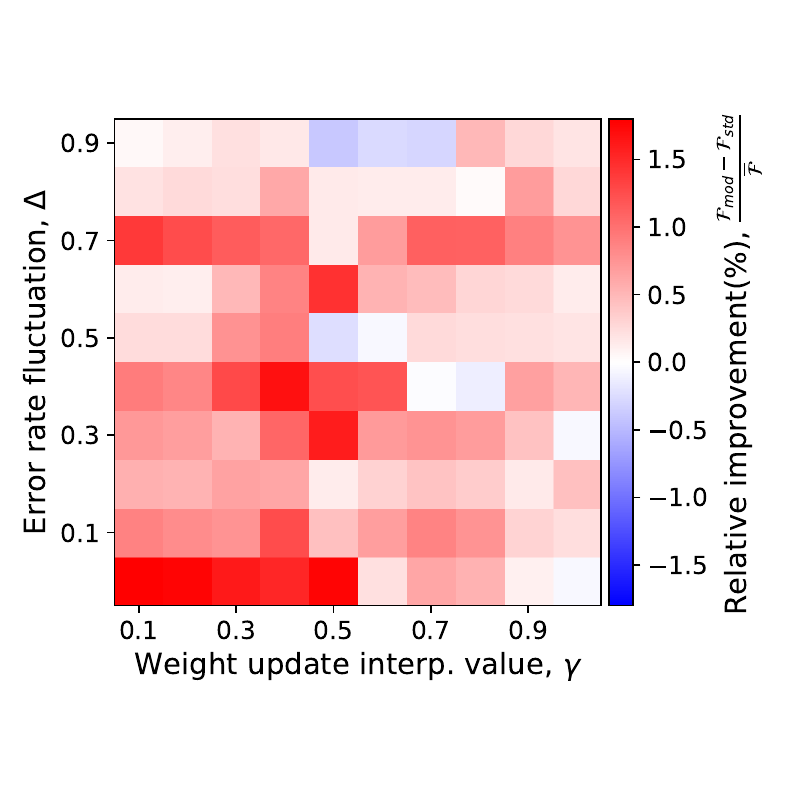}
        \caption{
            The relative improvement in fidelity of correlated MWPM compared to uncorrelated MWPM for simulated data with a heterogeneous error model. The scale of fluctuations of gate error rates $\Delta$ is shown on the vertical axis, while the strength of the weight update interpolation $\gamma$ is shown on the horizontal axis.
        }
        \label{fig:hetero_matrix}
    \end{figure}

    To investigate this further, we also simulate the surface-17 code under a heterogeneous error model where the error rates for two qubit gates and single qubit gates are drawn from normal distributions $\mathcal{N}(1.5\%,\Delta\cdot1\%)$ and $\mathcal{N}(0.09\%,\Delta\cdot0.04\%)$, respectively. The variable $\Delta$ scales the width of the normal distribution from which the error rates are drawn. The numerical values mimic now for $\Delta=1$ the noise level in the experiment (compare to Ref.~\onlinecite{Krinner2022}).
    Note that the error rate for each gate is only drawn once and then kept fix for all gate executions.
    The underlying idea for simulating a heterogeneous error model is that the resulting inferred weights will break degeneracies in the matching problem that would only be broken in correlated matching if one considered a homogenous error model where all error rates are identical.
    Following this line of reasoning, a more realistic and heterogeneous error model might reduce the improvement of the correlated matching.
    The parameter $\Delta$ can be used to control the degree of heterogeneity of the error model.
    In \cref{fig:hetero_matrix} the relative improvement is plotted for a fixed number of 16 stabilizer measurements by varied $\Delta$ and interpolation strength $\gamma$.
    The dataset size of this simulation is chosen as for the homogeneous error model simulation.
    Each value of $\Delta$ corresponds to a new set of simulated data.
    It is observable that for smaller $\Delta$, stronger updates should be performed.
    Overall, we recover the non-smooth behavior of the improvement metric, which we also found for the experimental data in \cref{fig:both_matrix} (a) as a result of increased statistical fluctuations.

        \section{Quantum Device}
    \label{sec:sup-s17-device}

    \begin{figure}
        \includegraphics[width=\linewidth]{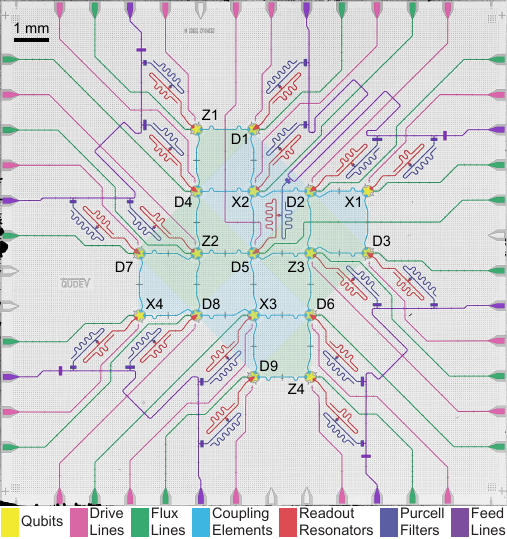}
        \caption{
            False-color optical micrograph of the 17-qubit quantum processor used in this work. See text for details about the device. Figure adapted from Ref.~\onlinecite{Krinner2022}.
        } \label{fig:device}
    \end{figure}

    The quantum processor is fabricated from a 150-nm-thin niobium film on a silicon substrate, from which the transmon islands, resonators and control lines are patterned using optical lithography.
    To realize coplanar waveguide crossovers and to connect different parts of the ground plane, we fabricate airbridges from an aluminum-titanium-aluminum trilayer.
    The Josephson junctions are fabricated from aluminum using electron beam lithography and shadow evaporation.
    Each of the 17 qubits (yellow) has an individual microwave drive line (pink) and flux line (green), see \cref{fig:device}.
    The qubits are capacitively coupled via a coplanar waveguides (cyan) to achieve an average qubit-qubit coupling rate of $J/2\pi\approx\SI{6}{MHz}$ and a mean interaction time for a dynamic flux pulse based \CZ{} gate~\cite{Negirneac2021} of~\SI{68}{ns}.
    For dispersive readout, we couple the qubits to readout resonators (red), which are coupled to four frequency-multiplexed feedlines (purple) via individual Purcell filters (blue) to suppress qubit decay and readout crosstalk~\cite{Heinsoo2018}.
    The qubit and resonator frequencies, anharmonicities and coherence properties are given in \cref{tab:qubit-parameters}.
    The readout resonator frequencies (\SI{6.769}{GHz} to \SI{7.554}{GHz}) are above the auxiliary qubits, which are biased to their upper flux-insensitive frequencies (\SI{5.885}{GHz} to \SI{6.192}{GHz}), and the data qubits biased to their lower flux-insensitive frequencies (\SI{3.740}{GHz} to \SI{4.143}{GHz}).
    The exception to the above is the auxiliary qubit \qb{X}{1}, which is biased to its lower flux-insensitive frequency to avoid interactions with a strongly coupled defect near its nominal bias frequency.
    We find a mean single qubit gate error of~\SI{0.09\pm0.04}{\percent} in randomized benchmarking~\cite{Magesan2011}, and a mean \CZ{} gate error of~\SI{1.5\pm1.0}{\percent} in interleaved randomized benchmarking~\cite{Magesan2012,Corcoles2013}.

 \begin{table*}
        \centering
        \caption{Qubit parameters, coherence properties and single-qubit performance for the nine data qubits (top) and the eight auxiliary qubits (bottom). We also provide, for relevant quantities, the averaged value across the device in column $\overline{\mathrm{Q}}$.}
        \begin{tabular}{lrrrrrrrrr}
            \toprule
            Parameter                                                      & D1    & D2    & D3    & D4    & D5    & D6    & D7    & D8    & D9                      \\
            \midrule
            Qubit idle frequency, $\omega_Q/2\pi$ (GHz)                    & 3.885 & 3.994 & 3.952 & 3.878 & 3.895 & 3.740 & 4.056 & 3.993 & 4.143                   \\
            Qubit anharmonicity, $\alpha/2\pi$ (MHz)                       & -184  & -183  & -183  & -184  & -186  & -184  & -181  & -183  & -181                    \\
            Lifetime, $T_1$ (\si{\micro s})                                & 31.1  & 29.0  & 69.9  & 55.5  & 32.7  & 59.1  & 33.2  & 25.8  & 30.3                    \\
            Ramsey decay time, $T_2^*$ (\si{\micro s})                     & 36.9  & 14.3  & 36.3  & 87.5  & 45.7  & 16.8  & 47.6  & 37.8  & 46.5                    \\
            Echo decay time, $T_2^\mathrm{e}$ (\si{\micro s})              & 47.2  & 48.6  & 46.5  & 87.8  & 54.2  & 24.1  & 49.9  & 45.9  & 51.7                    \\
            Readout frequency, $\omega_\mathrm{RO}/2\pi$ (GHz)             & 6.769 & 6.979 & 6.880 & 7.120 & 7.180 & 7.032 & 6.910 & 7.075 & 6.868                   \\
            Qb. freq. during RO, $\omega_Q'/2\pi$ (GHz)                    & 5.321 & 4.750 & 5.275 & 4.250 & 4.420 & 5.130 & 4.395 & 3.993 & 5.000                   \\
            Qubit-RO res. coupling, $g_{\mathrm{Q,RR}}/2\pi$ (MHz)         & 244   & 269   & 241   & 241   & 238   & 244   & 267   & 265   & 260                     \\
            Single-qubit RB error, $\epsilon_{\mathrm{1Q}}$ (\%)           & 0.06  & 0.07  & 0.04  & 0.04  & 0.06  & 0.06  & 0.04  & 0.08  & 0.06                    \\
            Two-state readout error, $\epsilon_{\mathrm{RO}}^{(2)}$ (\%)   & 0.7   & 0.6   & 0.5   & 2.7   & 1.9   & 1.0   & 2.0   & 0.8   & 0.4                     \\
            Three-state readout error, $\epsilon_{\mathrm{RO}}^{(3)}$ (\%) & 6.2   & 1.7   & 8.0   & 5.4   & 3.0   & 2.4   & 3.4   & 4.1   & 1.1                     \\
            \midrule
            Parameter                                                      & X1    & X2    & X3    & X4    & Z1    & Z2    & Z3    & Z4    & $\mathbf{\overline{Q}}$ \\
            \midrule
            Qubit idle frequency, $\omega_Q/2\pi$ (GHz)                    & 4.429 & 5.885 & 6.022 & 6.049 & 6.328 & 6.192 & 5.956 & 6.037 & \textbf{4.849}          \\
            Qubit anharmonicity, $\alpha/2\pi$ (MHz)                       & -181  & -174  & -170  & -170  & -163  & -168  & -171  & -170  & \textbf{-177}           \\
            Lifetime, $T_1$ (\si{\micro s})                                & 17.8  & 15.3  & 18.6  & 16.3  & 21.3  & 45.4  & 29.1  & 19.2  & \textbf{32.3}           \\
            Ramsey decay time, $T_2^*$ (\si{\micro s})                     & 21.6  & 20.3  & 21.9  & 27.8  & 37.8  & 34.2  & 49.8  & 25.8  & \textbf{35.8}           \\
            Echo decay time, $T_2^\mathrm{e}$ (\si{\micro s})              & 30.1  & 30.3  & 15.6  & 31.2  & 38.7  & 27.6  & 52.7  & 36.2  & \textbf{42.3}           \\
            Readout frequency, $\omega_\mathrm{RO/2\pi}$ (GHz)             & 7.372 & 7.554 & 7.258 & 7.461 & 7.316 & 7.502 & 7.200 & 7.412 & \textbf{7.170}          \\
            Qb. freq. during RO, $\omega_Q'/2\pi$ (GHz)                    & 5.900 & 5.885 & 6.022 & 6.049 & 6.328 & 6.192 & 5.956 & 6.037 & \textbf{5.347}          \\
            Qubit-RO res. coupling, $g_{\mathrm{Q-RR}}/2\pi$ (MHz)         & 167   & 168   & 167   & 168   & 171   & 170   & 167   & 171   & \textbf{213}            \\
            Single-qubit RB error, $\epsilon_{\mathrm{1Q}}$ (\%)           & 0.16  & 0.13  & 0.17  & 0.14  & 0.10  & 0.09  & 0.07  & 0.16  & \textbf{0.09}           \\
            Two-state readout error, $\epsilon_{\mathrm{RO}}^{(2)}$ (\%)   & 0.8   & 1.2   & 0.9   & 0.6   & 0.9   & 0.5   & 0.5   & 0.6   & \textbf{1.0}            \\
            Three-state readout error, $\epsilon_{\mathrm{RO}}^{(3)}$ (\%) & 1.3   & 3.2   & 3.5   & 1.5   & 2.1   & 1.6   & 5.2   & 1.3   & \textbf{3.2}            \\
            \bottomrule
        \end{tabular}
        \label{tab:qubit-parameters}
    \end{table*}

    \section{Correlations to Cycle Error Probabilities}
    \label{sec:sup-correlations-to-cycle-errors}

    We postulate that the probability $p_{i_1,\dots,i_n}$, that an error that flips $n$ syndrome elements $i_1$ to $i_n$ occurs, can be calculated from the observed correlations between the measured syndrome elements according to \cref{eq:corr-to-perr-general}, which we repeat here for the reader's convenience
    \begin{equation*}
        p_{i_1,\dots,i_n} = \frac{1}{2} - \frac{1}{2}\frac{
            \displaystyle\prod_{\begin{subarray}{r}
                    \{j_1,\dots,j_m\} \subseteq \\ \;\; \{{i_1,\dots,i_n}\}
                \end{subarray}}
            \left\langle \syndalt{j_1} \dots \syndalt{j_m} \right\rangle^{(-1)^{m-1} 2^{-(n-1)}}
        }{
            \displaystyle\prod_{\begin{subarray}{r}
                    \{j_1,\dots,j_m\} \supset \\ \;\; \{{i_1,...,i_n}\}
                \end{subarray}} \left(1 - 2 p_{j_1,\dots,j_m}\right)
        }.
        \tag{\ref{eq:corr-to-perr-general} repeated}
    \end{equation*}
    Again, we use the notation $\syndalt{i} = 1 - 2\synd{}{i} = \pm 1$ for the syndrome element at node $i$, for brevity.

    Assuming that there are no errors with signatures on more than two syndrome elements, we recover the formulas derived in~\cite{Spitz2018}
    \begin{subequations}
        \label{eq:corr-to-perr-spitz}
        \begin{align}
            p_{ij} {} & {}= \frac{1}{2} - \frac{1}{2}\sqrt{\frac{
                    \left\langle \syndalt{i} \right\rangle
                    \left\langle \syndalt{j} \right\rangle
                }{
                    \left\langle \syndalt{i} \syndalt{j} \right\rangle
                }} = \frac{1}{2} - \sqrt{\frac{1}{4} + \frac{
                    \left\langle \sigma_{i} \right\rangle \left\langle \sigma_{j} \right\rangle -  \left\langle \sigma_{i} \sigma_{j} \right\rangle
                }{
                    1 - 2 \left\langle \sigma_{i} \oplus \sigma_{j} \right\rangle
                }},
            \label{eq:corr-to-perr-spitz-double}                  \\
            p_{i} {}  & {}= \frac{1}{2} - \frac{1}{2} \frac{
                \left\langle \syndalt{i} \right\rangle
            }{
                \prod_{j\ne i} \left(1 - 2 p_{ij}\right)
            } = \frac{1}{2} - \frac{
                1/2 - \left\langle \sigma_{i} \right\rangle
            }{
                \prod_{j\ne i} \left(1 - 2 p_{ij}\right)
            },
            \label{eq:corr-to-perr-spitz-single}
        \end{align}
    \end{subequations}
    with $\oplus$ representing addition modulo two.

    Similarly, assuming no errors with signatures on more than four syndrome elements, we get the explicit formulas
    \begin{widetext}
        \begin{subequations}
            \begin{align}
                p_{ijkl} {} & {} = \frac{1}{2} - \frac{1}{2} \sqrt[8]{\frac{
                        \left\langle \syndalt{i} \right\rangle
                        \left\langle \syndalt{j} \right\rangle
                        \left\langle \syndalt{k} \right\rangle
                        \left\langle \syndalt{l} \right\rangle
                        \left\langle \syndalt{i} \syndalt{j} \syndalt{k} \right\rangle
                        \left\langle \syndalt{i} \syndalt{j} \syndalt{l} \right\rangle
                        \left\langle \syndalt{i} \syndalt{k} \syndalt{l} \right\rangle
                        \left\langle \syndalt{j} \syndalt{k} \syndalt{l} \right\rangle
                    }{
                        \left\langle \syndalt{i} \syndalt{j} \right\rangle
                        \left\langle \syndalt{i} \syndalt{k} \right\rangle
                        \left\langle \syndalt{i} \syndalt{l} \right\rangle
                        \left\langle \syndalt{j} \syndalt{k} \right\rangle
                        \left\langle \syndalt{j} \syndalt{l} \right\rangle
                        \left\langle \syndalt{k} \syndalt{l} \right\rangle
                        \left\langle \syndalt{i} \syndalt{j} \syndalt{k} \syndalt{l} \right\rangle
                }},                                                                                \\
                p_{ijk} {}  & {} = \frac{1}{2} - \frac{1}{2} \sqrt[4]{\frac{
                        \left\langle \syndalt{i} \right\rangle
                        \left\langle \syndalt{j} \right\rangle
                        \left\langle \syndalt{k} \right\rangle
                        \left\langle \syndalt{i} \syndalt{j} \syndalt{k} \right\rangle
                    }{
                        \left\langle \syndalt{i} \syndalt{j} \right\rangle
                        \left\langle \syndalt{i} \syndalt{k} \right\rangle
                        \left\langle \syndalt{j} \syndalt{k} \right\rangle
                }} \prod_{l\notin\{i,j,k\}} \frac{1}{1 - 2p_{ijkl}},                               \\
                p_{ij} {}   & {} = \frac{1}{2} - \frac{1}{2} \sqrt{\frac{
                        \left\langle \syndalt{i} \right\rangle
                        \left\langle \syndalt{j} \right\rangle
                    }{
                        \left\langle \syndalt{i} \syndalt{j} \right\rangle
                    }}
                \prod_{k\notin\{i,j\}} \left( \frac{1}{1 - 2p_{ijk}}
                \prod_{l\notin\{i,j,k\}} \frac{1}{1 - 2p_{ijkl}}
                \right),                                                                           \\
                p_{i} {}    & {}= \frac{1}{2} - \frac{1}{2} \left\langle \syndalt{i} \right\rangle
                \prod_{j\ne i} \left( \frac{1}{1 - 2p_{ij}}
                \prod_{k\notin\{i,j\}} \left( \frac{1}{1 - 2p_{ijk}}
                \prod_{l\notin\{i,j,k\}} \frac{1}{1 - 2p_{ijkl}}
                \right)\right).
            \end{align}
        \end{subequations}
    \end{widetext}

    We have explicitly derived \cref{eq:corr-to-perr-general} for errors triggering up to six syndromes by solving the equations
    \begin{align}
        \label{eq:equations-to-solve}
        \syndalt{i_1} = {} & {} F_{i_1}
        \prod_{i_2 \ne i_1} \Biggl( F_{i_1,i_2}
        \prod_{i_3 \notin \{i_1, i_2\} } \Biggl( F_{i_1,i_2,i_3}
        \nonumber                       \\ {}&{}
        \prod_{i_4 \notin \{i_1, i_2, i_3\} } \Biggl( F_{i_1,i_2,i_3,i_4}
        \prod_{i_5 \notin \{i_1, i_2, i_3, i_4\} } \Biggl( F_{i_1,i_2,i_3,i_4,i_5}
        \nonumber                       \\ {}&{}
        \prod_{i_6 \notin \{i_1, i_2, i_3, i_4, i_5\} } F_{i_1,i_2,i_3,i_4,i_5,i_6} \Biggr)\Biggr)\Biggr)\Biggr),
    \end{align}
    where $F_{i_1,\dots,i_n} = \pm 1$ represents the underlying random variable that indicates whether the error that flips syndrome elements $\{i_1,\dots,i_n\}$ has happened ($-1$) or not ($+1$).
    The value of $F_{i_1,\dots,i_n}$ does not depend on the order of its indices, and no two indices should have the same value.
    We further assume that the error processes are independent, that is $\langle F_{i_1,\dots,i_n} F_{j_1,\dots,j_m} \rangle = \langle F_{i_1,\dots,i_n} \rangle \langle F_{j_1,\dots,j_m} \rangle$ if $\{i_1,\dots,i_n\} \ne \{j_1,\dots,j_m\}$, and that the probability of each error happening is given by $p_{i_1,\dots,i_n} = (1 - \langle F_{i_1,\dots,i_n} \rangle)/2$.

\begin{figure}[t]
        \includegraphics[width=\columnwidth]{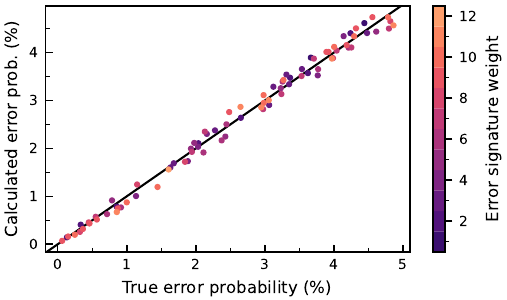}
        \caption{Error probability as calculated according to~\cref{eq:corr-to-perr-general} as a function of the true simulated error probability for 83 random error channels with signatures on up to 12 nodes. The simulation was done with 100000 shots and took about 30 minutes on a laptop computer.
        } \label{fig:method-numerical-verif}
    \end{figure}

    We also numerically validated \cref{eq:corr-to-perr-general} on artificial datasets, where correlated errors triggering up to twelve syndromes exist, see \cref{fig:method-numerical-verif}.

    \section{Bias in Cycle Error Calculation}
    \label{sec:sup-cycle-error-bias}

    We present an example of how not accounting for highly correlated error signatures can bias the extracted probabilities of errors with lower-weight signatures.
    Let us consider three syndrome elements $\sigma_1$, $\sigma_2$, and $\sigma_3$, which are affected by three independent error processes.
    The first process flips the syndrome element $\sigma_1$ with probability $p_1=\SI{3}{\percent}$, the second flips both syndrome elements $\sigma_1$ and $\sigma_2$ with probability $p_{12}=\SI{2.5}{\percent}$, and the third process flips all syndrome elements with probability $p_{123}=\SI{1}{\percent}$.
    The expectation values for the correlations between syndrome elements are given in \cref{tab:correlation-expectations}.
    If we use \cref{eq:corr-to-perr-general} to calculate the error probabilities from the correlations, we recover the original probabilities $p_1$, $p_{12}$ and $p_{123}$.
    However, if we use the simplified formula \cref{eq:corr-to-perr-spitz} that does not account for the highly correlated error process $p_{123}$, we get erroneous probabilities $\check{p}$, see \cref{tab:inferred-probs}.
    We see that two-way correlations are overestimated by roughly $p_{123}$, because the three-way error process flips all pairs of syndromes.
    However, the single-syndrome-flipping error probabilities are underestimated by $p_{123}$, because the two-way error probabilities that are effectively subtracted by the denominator in \cref{eq:corr-to-perr-spitz-single} were overestimated.
    We see that in some cases ($\check{p}_2$ and $\check{p}_3$), the extracted probabilities can even appear negative if some error channels are not accounted for.

    \begin{table}[t]
        \centering
        \caption{Syndrome correlations for the three-node example.}
        \label{tab:correlation-expectations}
        \begin{tabular}{llr}
            \toprule
            Correlation                                           & Equation                                & \multicolumn{1}{l}{Num.} \\
            \midrule
            $\langle \syndalt{1} \rangle$                         & $(1 - 2p_1)(1 - 2p_{12})(1 - 2p_{123})$ & $0.875$                  \\
            $\langle \syndalt{2} \rangle$                         & $(1 - 2p_{12})(1 - 2p_{123})$           & $0.931$                  \\
            $\langle \syndalt{3} \rangle$                         & $1 - 2p_{123}$                          & $0.980$                  \\
            $\langle \syndalt{1} \syndalt{2} \rangle$             & $1 - 2p_1$                              & $0.940$                  \\
            $\langle \syndalt{2} \syndalt{3} \rangle$             & $1 - 2p_{12}$                           & $0.950$                  \\
            $\langle \syndalt{1} \syndalt{3} \rangle$             & $(1 - 2p_1)(1 - 2p_{12})$ & $0.893$                  \\
            $\langle \syndalt{1} \syndalt{2} \syndalt{3} \rangle$ & $(1 - 2p_1)(1 - 2p_{123})$              & $0.921$                  \\
            \bottomrule
        \end{tabular}
    \end{table}

    \begin{table}[t]
        \centering
        \caption{Error probabilities $\check{p}$ inferred using \cref{eq:corr-to-perr-spitz} for the three-node example.}
        \label{tab:inferred-probs}
        \begin{tabular}{llrr}
            \toprule
            Inferred         &                                 &                          &                           \\
            probability      & Equation                        & \multicolumn{1}{l}{Num.} & \multicolumn{1}{l}{Error} \\
            \midrule
            $\check{p}_1$    & $(p_1 - p_{123})/(1-2p_{123})$  & $0.020$                  & $-0.010$                  \\
            $\check{p}_2$    & $-p_{123}/(1-2p_{123})$         & $-0.010$                 & $-0.010$                  \\
            $\check{p}_3$    & $-p_{123}/(1-2p_{123})$         & $-0.010$                 & $-0.010$                  \\
            $\check{p}_{12}$ & $p_{12}+p_{123}-2p_{12}p_{123}$ & $0.034$                  & $0.010$                   \\
            $\check{p}_{23}$ & $p_{123}$                       & $0.010$                  & $0.010$                   \\
            $\check{p}_{13}$ & $p_{123}$                       & $0.010$                  & $0.010$                   \\
            \bottomrule
        \end{tabular}
    \end{table}

    \section{Simulation Model}
    \label{sec:sup-pauli-simulation-model}

    To obtain the simulation data shown in \cref{fig:3}~(b), we conduct a Clifford simulation using the PECOS~\cite{Ryan-Anderson2018} package, where we implement the circuit presented in \cref{sec:sup-full-circuit-diagram}.
    Single- and two-qubit depolarizing noise is implemented by applying a random single- or two-qubit Pauli operator to the state with probability $p$.
    We use a uniform noise model, where operations on each qubit are subject to the same error probabilities.
    We apply single-qubit depolarizing noise during each idling step of duration $t_\mathrm{IDL}$ with probability $p_\mathrm{IDL} = (1-e^{-t_\mathrm{IDL}/\overline{T}})/4$, where $\overline{T} = \SI{35}{\micro s}$ is approximately the mean $T_1$ and $T_2$ time of all the qubits.
    For single-qubit gates, we apply a single-qubit depolarizing channel with probability $p_\mathrm{1Q} = 0.0009$.
    That is the average of single-qubit gate errors from randomized benchmarking.
    Similarly, for two-qubit gates, we apply a two-qubit depolarizing channel with probability $p_\mathrm{2Q} = 0.015$, the average gate error from interleaved randomized benchmarking, see \cref{sec:sup-s17-device}.
    Finally, to model readout errors, we apply a $\sigx$ gate before each readout with probability $p_\mathrm{RO} = 0.0116$, which corresponds approximately to the average two-state readout error $\epsilon_\mathrm{RO}^{(2)}$, see \cref{tab:qubit-parameters}.

    \section{Full Circuit Diagram}
    \label{sec:sup-full-circuit-diagram}

    \begin{figure*}
        \includegraphics[width=\linewidth]{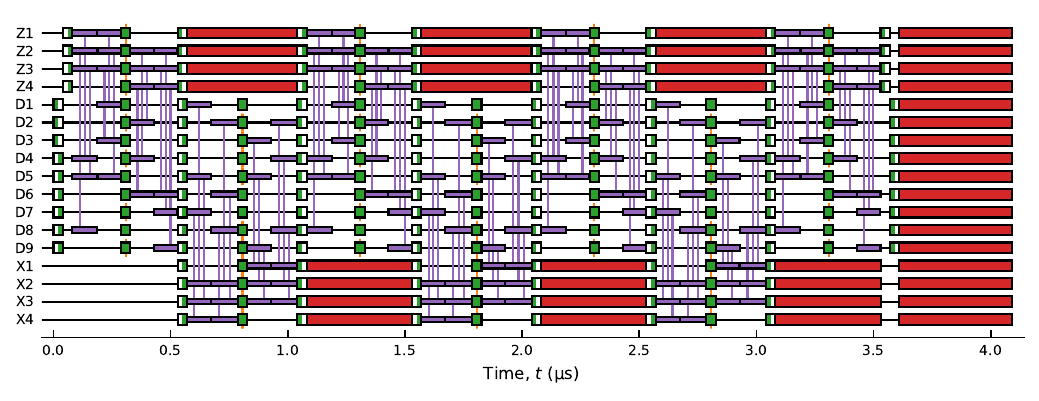}
        \caption{
            Circuit diagram for a four-cycle-long logical state preservation experiment for the $\ket{+}_\logical$ state.
            Green boxes denote single qubit rotations around $\sigy$, with the filling indicating the rotation angle: left half filled is $-\pi/2$, right half filled is $\pi/2$, and fully filled is $\pi$. Orange vertical lines indicate virtual $\pi$-rotations around $\sigz$, which are implemented by flipping the phase of all following single-qubit gates.
            Red boxes indicate readout and pairs of purple boxes connected by a vertical line denote \CZ{} gates.
        } \label{fig:operations}
    \end{figure*}

    A diagram of an example circuit four-cycle-long logical state preservation experiment for the $\ket{+}_\logical$ state is shown in \cref{fig:operations}.
    The experimental data presented in this work was acquired using an equivalent 16-cycle-long circuit, preparing either $\ket{0}_\logical$, $\ket{1}_\logical$, $\ket{+}_\logical$, or $\ket{-}_\logical$.
    During the first stabilizer measurement cycle, we omit measuring the X-type (Z-type) stabilizers when preparing an eigenstate of the $\XL$ ($\ZL$) operator.

    \section{Example analysis of varying error rates}
    \label{sec:sup-time-drift-example}

    As the simplest example that shows how the changes in the underlying error rates of the system during the data aquisition time can lead to apparent correlated errors, we consider the following system. There are two syndrome elements which independently flip with some probability. In the first scenario, let the error probability be constantly $p$ throughout the data-gathering time. In this case, the syndrome correlations are
    \begin{subequations}
        \begin{align}
            \langle\syndalt{1}\rangle = \langle\syndalt{2}\rangle & = 1 - 2p,     \\
            \langle\syndalt{1} \syndalt{2}\rangle                 & = (1 - 2p)^2,
        \end{align}
    \end{subequations}
    and the error probabilities extracted using \cref{eq:corr-to-perr-general} are
    \begin{subequations}
        \begin{align}
            p_{12} & = \frac{1}{2} - \frac{1}{2} \sqrt{\frac{\langle\syndalt{1}\rangle \langle\syndalt{2}\rangle}{\langle\syndalt{1}\syndalt{2}\rangle}} = 0, \\
            p_1    & = \frac{1}{2} - \frac{1}{2}  \langle\syndalt{1}\rangle = p,                                                                              \\
            p_2    & = \frac{1}{2} - \frac{1}{2}  \langle\syndalt{2}\rangle = p.
        \end{align}
    \end{subequations}

    In a second scenario, we set the the individual error probabilities to zero for the first half of the data-gathering and to $2p$ for the second half, yielding syndrome correlations
    \begin{subequations}
        \begin{align}
            \langle\syndalt{1}\rangle = \langle\syndalt{2}\rangle & = 1/2 + (1 - 4p)/2 = 1 - 2p,              \\
            \langle\syndalt{1} \syndalt{2}\rangle                 & = 1/2 + (1 - 4p)^2/2 = (1 - 2p)^2 + 4p^2.
        \end{align}
    \end{subequations}
    We now find a nonzero probability for an error simultaneously flipping the two syndrome elements:
    \begin{subequations}
        \begin{align}
            p_{12} ={} & {} \frac{1}{2} - \frac{1}{2} \sqrt{\frac{\langle\syndalt{1}\rangle \langle\syndalt{2}\rangle}{\langle\syndalt{1}\syndalt{2}\rangle}} = \notag \\
                       & {}\frac{1}{2} - \frac{1}{2} \sqrt{\frac{(1-2p)^2}{(1 - 2p)^2 + 4p^2}} \ne 0,                                                                  \\
            p_1 ={}    & {} \frac{1}{2} - \frac{1}{2}  \frac{\langle\syndalt{1}\rangle }{1-2p_{12}} \ne p,                                                             \\
            p_2 ={}    & {} \frac{1}{2} - \frac{1}{2}  \frac{\langle\syndalt{2}\rangle }{1-2p_{12}} \ne p.
        \end{align}
    \end{subequations}
    In this extreme example, where the errors are fully off for half of the experiment, the apparent correlated error probability is equal to $p^2$ in the limit of small $p$.
    For a smaller change in the error rates of $\varepsilon$, that is, error probabilities $(1-\varepsilon)p$ for half of the data gathering time and $(1+\varepsilon)p$ for the second half, we would find a correlated error probability proportional to $\varepsilon^2 p^2$.
    Due to the quadratic scaling in both $\varepsilon$ and $p$, the effect is significant only for large changes in the error rates.

\end{appendix}

\end{document}